%% file: paper.tex
\documentclass[preprint,journal]{vgtc}            

\onlineid{1463}

\vgtccategory{Research}

\vgtcpapertype{algorithm/technique}

\usepackage[utf8]{inputenc}
\usepackage[svgnames]{xcolor}
\usepackage[pagebackref,bookmarks]{hyperref}
\usepackage{tikz}
\usetikzlibrary{shapes.geometric}
\usepackage{amssymb}
\usepackage{amsmath}
\usepackage{mathtools}
\usepackage[disable]{todonotes}
\usepackage[linesnumbered,noend,ruled]{algorithm2e}
\usepackage{enumitem}
\usepackage{appendix}
\usepackage{amssymb}
\usepackage{pifont}
\usepackage{amsthm}
\usepackage{float}
\usepackage{doi}
\usepackage[export]{adjustbox}
\usepackage{filecontents}

\usepackage{xr-hyper}
\makeatletter
\newcommand*{\addFileDependency}[1]{
  \typeout{(#1)}
  \@addtofilelist{#1}
  \IfFileExists{#1}{}{\typeout{No file #1.}}
}
\makeatother

\newcommand*{\myexternaldocument}[1]{%
    \externaldocument{#1}%
    \addFileDependency{#1.tex}%
    \addFileDependency{#1.aux}%
}

\myexternaldocument{supp}

\DeclareMathOperator{\troot}{root}
\DeclareMathOperator{\edges}{edges}

\DeclareMathOperator{\cost}{c}
\DeclareMathOperator{\pathstart}{start}
\DeclareMathOperator{\pathend}{end}

\DeclareMathOperator{\mappingedit}{edits}
\DeclareMathOperator{\mappingrelabel}{rel}
\DeclareMathOperator{\mappinginsert}{ins}
\DeclareMathOperator{\mappingdelete}{del}
\newcommand{\pmdist}{\delta_{\text{P}}}

\newtheorem{definition}{Definition}

\colorlet{MarkColor}{black}

\title{Merge Tree Geodesics and Barycenters with Path Mappings}

\author{%
  \authororcid{Florian Wetzels}{0000-0002-5526-7138},
  \authororcid{Mathieu Pont}{0000-0002-0037-0314},
  \authororcid{Julien Tierny}{0000-0003-0056-2831}, and
  \authororcid{Christoph Garth}{0000-0003-1669-8549}
}

\authorfooter{
  \item
  	Florian Wetzels and Christoph Garth are with University of Kaiserslautern-Landau.
  	E-mail: \{wetzels\,$|$\,garth\}@rptu.de
  \item
  	Mathieu Pont and Julien Tierny are with CNRS / Sorbonne University.
  	E-mail: \{mathieu.pont\,$|$\,julien.tierny\}@sorbonne-universite.fr
}

\abstract{Comparative visualization of scalar fields is often facilitated using similarity measures such as edit distances.
In this paper, we describe a novel approach for similarity analysis of scalar fields that combines two recently introduced techniques: Wasserstein geodesics/barycenters
as well as path mappings, a branch decomposition-independent edit distance.
Effectively, we are able to leverage the reduced susceptibility of path mappings to small perturbations in the data when compared with the original Wasserstein distance.
Our approach therefore exhibits superior performance and quality in typical tasks such as ensemble summarization, ensemble clustering, and temporal reduction of time
series, while retaining practically feasible runtimes.
Beyond studying theoretical properties of our approach and discussing implementation aspects, we describe a number of case studies that provide empirical insights into its utility for comparative visualization, and demonstrate the advantages of our method in both synthetic and real-world scenarios.
We supply a C++ implementation that can be used to reproduce our results.%
}


\teaser{
  \centering
  \includegraphics[angle=270,width=160mm]{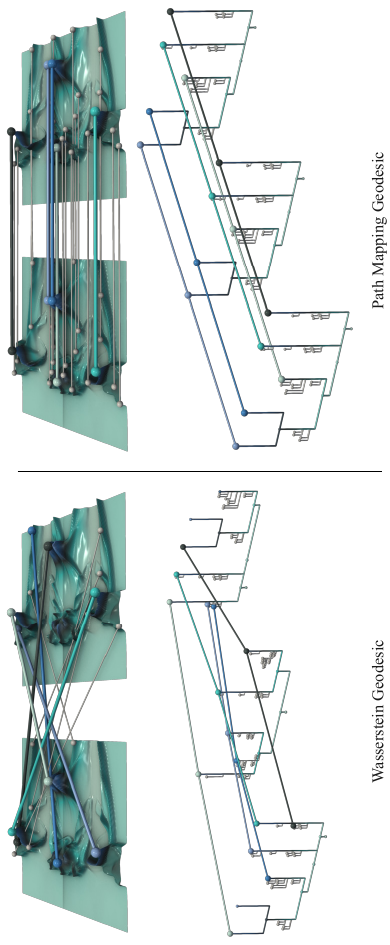}
  \caption{Two merge tree geodesics with their corresponding mappings.
  The top row shows the mappings of critical points (restricted to maxima) embedded into the domain.
  In the bottom row the mapping between the two \textcolor{MarkColor}{corresponding} merge trees and their interpolated geodesic is shown.
  The most persistent mapped features are highlighted and colored to show the correspondence between embedding and planar layout.
  The Wasserstein geodesic can be seen on the left, the path mapping geodesic on the right.
  Clearly, the middle tree on the right is a more meaningful interpolation of the two input trees than the middle tree on the left; and yields a more intuitive correspondence between the nodes.
  Note that the used layout differs between the two figures.
  However, the  two trees in front are indeed identical in both figures, as are the trees in the back.}
  \label{fig:teaser}
}

\graphicspath{{figs/}{figures/}{pictures/}{images/}{./}} 

\usepackage{tabu}                      
\usepackage{booktabs}                  
\usepackage{lipsum}                    
\usepackage{mwe}                       

\usepackage{mathptmx}                  

\begin{document}


\firstsection{Introduction}
\label{sec:intro}

\maketitle


The comparison of scalar fields through edit distances or related similarity measures is an area of increasing interest in scientific visualization.
Specifically in the growing field of ensemble visualization, comparative analysis techniques are of high interest, as the data becomes more and more complex.
Since the size of scalar fields acquired from real-world experiments or simulation quickly makes direct comparison on the domains infeasible, topological abstractions such as merge trees, contour trees or persistence diagrams are used to derive efficient distance measures.
This has lead to a rapid development in comparison techniques on merge trees in recent years~\cite{SaikiaSW14_branch_decomposition_comparison,BeketayevYMWH14,morozov14,SridharamurthyM20,pont_vis21,DBLP:journals/cgf/WetzelsLG22,path_mappings}.

Out of those, two prominent examples are Wasserstein barycenters for merge trees and branch decomposition-independent edit distances.
The Wasserstein barycenter framework by Pont et al.~\cite{pont_vis21} introduced new analysis techniques based on barycenters and geodesics, in regards to the Wasserstein distance for merge trees.
\textcolor{MarkColor}{Intuitively, barycenter merge trees are an interpolated mean for a collection of merge trees, based on a given metric, while a geodesic is a shortest continuous path in this metric space of trees.}
The distance is an extension of the well-known Wasserstein distance for persistence diagrams and is based on the so-called degree-1 edit distance on branch decomposition trees (BDT).
Applications are, e.g., k-means~\cite{elkan03,celebi13} clustering or temporal reduction of time series.
In contrast to that, the work of \textcolor{MarkColor}{Wetzels et al.\ introduced} two novel comparison measures for merge trees: the branch mapping distance~\textcolor{MarkColor}{\cite{DBLP:journals/cgf/WetzelsLG22}} and the path mapping distance~\textcolor{MarkColor}{\cite{path_mappings}.}
These are edit distances tailored specifically to merge trees yielding better quality comparison at the cost of increased complexity.
The branch mapping distance introduced the concept of branch decomposition-independent edit distances, however, it does not fulfill the metric properties.
The path mapping distance is an improved variant which forms a metric and represents an intuitive edit distance based on deformation retractions on merge trees as continuous topological spaces.

In this paper, we propose a combination of these techniques such that the merge tree barycenter framework can take advantage of the improved distance measures.
More specifically, we implement barycenters and geodesics based on the path mapping distance, since it is a metric, making it better suited than the branch mapping distance.
We thereby increase the flexibility of the framework: depending on the type, size and structure of the data, it is suited for high quality analysis using the more precise but also more complex path mapping distance as well as efficient and less precise analysis based on Wasserstein distances.
We then evaluate the new method in terms of quality and performance on synthetic and real-world datasets.
For the evaluation, we used three visualizations and analysis tasks on which barycenters and geodesics can be applied: ensemble summarization, ensemble clustering and temporal reduction of time series.
An example for significantly improved interpolation on merge trees is illustrated in \autoref{fig:teaser}.
We also briefly compare path mapping barycenters to the only other method we know of computing a merge tree representing a whole ensemble: the contour tree alignment~\cite{LohfinkWLWG20}.
Here, we illustrate the fact the barycenter merge trees are indeed valid merge trees, whereas (in general) the alignment is not.
In particular, our contributions are:
\begin{itemize}
    \item We describe an algorithm to compute both path mapping geodesics and path mapping barycenters.
    \item We provide an experimental evaluation for the utility of the new
    method. Our experiments are -- to our knowledge -- the first that show advantages of branch
    decomposition-independent comparison of real-world datasets, and advantages of path mappings over
    branch mappings. Previously, the former had only been shown on synthetic
    examples~\cite{DBLP:journals/cgf/WetzelsLG22,path_mappings}).
    \item We provide an open source implementation that is based on the \emph{Topology ToolKit} (TTK)~\cite{DBLP:journals/tvcg/TiernyFLGM18}, extending its barycenter framework.
\end{itemize}

\noindent
We conclude this section with an overview over related work. In \autoref{sec:preliminaries},
we give the formal background and recap path mappings and the Wasserstein barycenter framework.
\autoref{sec:method} describes the new algorithm for path mapping barycenters and geodesics.
In \autoref{sec:experiments}, we present the results of our experiments, before \autoref{sec:conclusion} concludes the paper.

\subsection*{Related Work}

Topological abstraction or descriptors for scalar fields are a well-established tool in scientific visualization with applications covering a vast area of tasks such as assisting rendering and interaction~\cite{TAKAHASHI200424,pascucci_mr04,DBLP:conf/vissym/CarrS03}, comparing data~\cite{surveyComparison2021} or deriving abstract visualizations~\cite{DBLP:journals/tvcg/WeberBP07,LohfinkWLWG20}.
An overview over these methods can be found in the survey by Heine et al.~\cite{heine16}.
We identified three areas specifically related to our work: topology-based comparison of scalar fields, feature tracking (or more general analysis of time series) and ensemble visualization.

\textbf{Similarity Measures.}
Scalar field comparison through distances on merge trees include several edit distances by Saikia et al.~\cite{SaikiaSW14_branch_decomposition_comparison}, Sridharamurthy et al.~\cite{SridharamurthyM20,DBLP:journals/tvcg/SridharamurthyN23}, Pont et al.~\cite{pont_vis21} and Wetzels et al.~\cite{DBLP:journals/cgf/WetzelsLG22,path_mappings}.
Examples for other merge tree-based distances are the works of Beketayev et al.~\cite{BeketayevYMWH14}, Morozov et al.~\cite{MorozovW14} or Bollen et al~\cite{DBLP:journals/tvcg/BollenTL23}.

Apart from merge trees, scalar field comparison is often done via distances on other topological descriptors such as contour trees~\cite{DBLP:journals/tvcg/ThomasN11,LohfinkWLWG20}, persistence diagrams~\cite{edelsbrunner09,Cohen-Steiner2010,Cohen-Steiner2007,interleaving_distance}, reeb graphs~\cite{bauer14,DBLP:journals/dcg/FabioL16,DBLP:conf/3dor/BauerFL16} or extremum graphs~\cite{ThomasN13,DBLP:conf/apvis/NarayananTN15}.
Some distances also combine geometric and topological measures~\cite{intrinsicMTdistance,YanWMGW20,9744472}.
A survey on the methods can be found in~\cite{surveyComparison2021}.

\textbf{Feature Tracking and Time Series.}
We classify topological methods for feature tracking into two major categories.
One uses mappings between topological descriptors to derive mappings between the features of consecutive time steps.
Examples based on merge trees are the works of Lohfink et al.~\cite{DBLP:journals/corr/abs-2107-12682}, Saikia et al.~\cite{SaikiaSW14_branch_decomposition_comparison} or Pont et al.~\cite{pont_vis21}, while other descriptors are used as well~\cite{oesterling2017computing,DBLP:conf/compgeom/EdelsbrunnerHMP04}.
The other class of methods uses topological descriptors for feature identification and then applies geometric techniques such as gradient or overlap mapping to find correspondences.
Examples are the works of Lukasczyk et al.~\cite{LukasczykGWBML20,lukasczyk2017,DBLP:journals/cgf/LukasczykWMGL17}, Bremer et al.~\cite{bremer_tvcg11,bremer2009analyzing,widanagamaachchi2012interactive} or others~\cite{SaikiaW17,DBLP:conf/apvis/ShuGLCLY16,nilsson2022exploring,DBLP:journals/tvcg/SchnorrHDKH20}.
A method by Yan et al.~\cite{9744472} combines geometric and topological
measures.
Furthermore, topological methods can also be used for more advanced
analysis of time series such as temporal merge tree
maps~\cite{DBLP:journals/tvcg/KoppW23} or
geodesics~\cite{Turner2014,pont_vis21}.

\textbf{Ensemble and Uncertainty Visualization.}
Topological methods for ensemble analysis include the above mentioned similarity measures on topological descriptors.
Apart from that, a typical approach is, given an ensemble of topological descriptors, to compute a representative summarizing the set of member descriptors.
Examples are fuzzy contour trees~\cite{LohfinkWLWG20},
merge tree 1-centers~\cite{YanWMGW20},
the uncertain contour tree layout~\cite{Wu2013ACT},
coherent contour trees~\cite{DBLP:conf/grapp/EversHML20},
and Wasserstein barycenters of merge trees~\cite{pont_vis21} or persistence diagrams~\cite{vidal_vis19}.
Based on the merge tree barycenters, advanced statistical tools have also been proposed, such as principal geodesic analysis for merge trees~\cite{pont_tvcg23}.
Other methods for ensemble or uncertain data are found in~\cite{Wu2013ACT,Kraus2010VisualizationOU,gunther,Turner2014}.

\section{Preliminaries}
\label{sec:preliminaries}

In this section, we quickly recap the concepts and definitions of merge trees, the path mapping distance and the Wasserstein barycenters.
In most cases, we simply re-state the definitions from~\cite{path_mappings,DBLP:journals/cgf/WetzelsLG22,pont_vis21}.

\subsection{Merge Trees}

Given a domain $\mathbb{X}$ with a continuous scalar function  $f: \mathbb{X} \rightarrow \mathbb{R}$, the \emph{merge trees} of $\mathbb{X},f$ represent the connectivity of sublevel or superlevel sets.
Depending on the
direction, we call them \emph{join tree} or \emph{split tree};
both form rooted trees.
In this paper, we work on discrete representations of these rooted trees, called abstract merge trees and defined in the following.
The nodes of an abstract merge tree are critical points of the domain $\mathbb{X},f$ and the edges represent classes of connected components of level sets.
For simplicity and w.l.o.g.\ we restrict to abstract split trees in the theoretic discussions (all arguments and definitions are easy to adapt for join trees).
A detailed introduction is given in \cite{DBLP:conf/focs/EdelsbrunnerLZ00,DBLP:conf/ppopp/MorozovW13}.

A rooted, unordered tree $T$ is a connected, directed graph with vertex set $V(T)$ and edge set $E(T)$, containing no undirected cycles and a unique sink, which we call the root, denoted $\troot(T)$.
For an edge $(c,p) \in E(T)$, we call $c$ the child of $p$ and $p$ the parent of $c$, i.e.\ we use parent pointers in our directed representation.

As for general graphs, a \emph{path} of length $k$ in a rooted tree $T$ is a sequence of vertices $p=v_1 \dots v_k \in V(T)^k$ with $(v_{i},v_{i-1}) \in E(T)$ for all $2 \leq i \leq k$ and $v_i \neq v_j$ for all $1 \leq i < j \leq k$. Note the strict root-to-leaf direction of the vertex sequence: we only consider monotone paths.
For a path $p=v_1 \dots v_k$, we denote its first vertex by $\pathstart(p) := v_1$, its last vertex by $\pathend(p) := v_k$.
We write $v \in p$ if $v = v_i$ for some $1 \leq i \leq k$ and $e \in p$ if $e = (v_i,v_{i-1})$ for some $2 \leq i \leq k$.
We denote the set of edges of $p$ by $\edges(p) := \{(v_{i},v_{i-1})\}$.
We denote the set of all paths of a tree $T$ by $\mathcal{P}(T)$.

Merge trees inherit the scalar function from their original domain and thus are labeled trees.
Usually, they are considered as node-labeled trees, with the labels being defined through the scalar function on the critical points.
All merge trees have certain properties in common, which we now use for the definition of an abstract merge tree.
\begin{definition}
An unordered, rooted tree $T$ of (in general) arbitrary degree (i.e.\ number of children) with node labels $f:V(T) \rightarrow \mathbb{R}_{>0}$ is an \emph{Abstract Merge Tree} if the following properties hold:
\begin{itemize}
    \item the root node has degree one, $\deg( \troot (T) ) = 1$,
    \item all inner nodes have a degree of at least two,\\ $\deg(v) \neq 1$ for all $v \in V(T)$ with $v \neq \troot (T)$,
    \item all nodes have a larger scalar value than their parent node, $f(c) > f(p)$ for all $(c,p) \in E(T)$.
\end{itemize}
\end{definition}
\noindent
For the path mapping distance and the corresponding geodesics, we work on edge-labeled trees.
Here, edge labels represent the length of the scalar range of the edge.
These two representations are interchangeable;
given a node label function $f : V(T) \rightarrow \mathbb{R}_{>0}$, we define the corresponding edge label function $\ell_f : E(T) \rightarrow \mathbb{R}_{>0}$ by $\ell_f((u,v)) = |f(u) - f(v)|$.
Given an edge label function, we can again define $f_\ell$ by placing the root node at a fixed scalar value, e.g.~$0$.

We lift the edge label function $\ell$ of a merge tree $T$ from edges to paths in the following way: $\ell(v_1 \dots v_k) = \sum_{2 \leq i \leq k} \ell((v_i,v_{i-1}))$.
For the implicit edge labels $\ell_f$, we get $\ell_f(v_1 \dots v_k) = |f(v_1) - f(v_k)|$.
To define edit distances on merge trees, we also need to define a cost function to compare edges.
Since we use $\mathbb{R}_{>0}$ as the label set for abstract merge trees, we use~$0$ as the blank symbol, i.e.\ the label of an empty or non-existing edge.
Then, we define the cost function as the euclidean distance on $\mathbb{R}_{\geq 0}$: $ \cost(l_1,l_2) = |l_1-l_2| $ for all $l_1,l_2 \in \mathbb{R}_{\geq 0}$.

A \emph{branch} of an abstract merge tree $T$ is a path that ends in a leaf.
A branch $b=b_1 ... b_k$ is a parent branch of another branch $a=a_1 ... a_\ell$ if $a_1 = b_i$ for some $1 < i < k$.
We also say $a$ is a child branch of $b$.
A set of branches $B=\{B_1,...,B_k\}$ of a merge tree $T$ is called a \emph{Branch Decomposition} of $T$ if $\{\edges(B_1),...,\edges(B_k)\}$ is a partition of $E(T)$.
We also call the length of a branch its \emph{persistence}.
The parent-child relations of the branches in a branch decomposition $B$ of $T$ form a tree structure by themselves.
The tree built from the vertex set $V=B$ and edge set $E$, with $(a,b) \in E$ if and only if $b$ is a parent branch of $a$, is called the \emph{Branch Decomposition Tree} (BDT) of $B$.
In practice, the nodes of the BDT are labeled with the the birth and death values of its branch (i.e. the scalar values of the first and last vertex~\cite{edelsbrunner09}).
Most of the time, the branch decomposition derived by the elder rule (giving longer/more persistent branches higher priority, see~\cite{edelsbrunner09} for details) is used.
We denote it by $\mathcal{B}(\mathbb{X},f)$ for a scalar field $\mathbb{X},f$.
For simplicity, we often write $b \in \mathcal{B}(\mathbb{X},f)$ instead of $b \in V(\mathcal{B}(\mathbb{X},f))$.

\subsection{Path Mapping Distance}

We now recap the concepts of \emph{path mappings} and the \emph{path mapping distance} between merge trees.
Both were introduced in~\cite{path_mappings} to capture a constrained variant of the deformation based edit distance defined in the same paper.
In contrast to classic edit mappings that map the nodes of two trees onto each other, path mappings (as the name suggests) map paths of one tree to paths of another tree.
Similar to classic edit mappings, they do so in a structure-preserving way.
We now re-state the definition of path mappings and the corresponding distance measure.

\begin{definition}
\label{definiton:path_mappings}

Given two abstract merge trees $T_1,T_2$, a path mapping between $T_1$ and $T_2$ is a mapping $M \subseteq \mathcal{P}(T_1) \times \mathcal{P}(T_2)$ such that
\begin{enumerate}
    \item \textcolor{MarkColor}{(one-to-one)} for all $p_1,q_1 \in \mathcal{P}(T_1)$, $p_2,q_2 \in \mathcal{P}(T_2)$ with $(p_1,p_2) \in M$ and $(q_1,q_2) \in M$, $p_1=q_1$ if and only if $p_2=q_2$,
    \item \textcolor{MarkColor}{(paths do not overlap)} $|p_1 \cap q_1| \leq 1$ and $|p_2 \cap q_2| \leq 1$ for all $(p_1,p_2),(q_1,q_2) \in M$,
    \item \textcolor{MarkColor}{(paths form a connected subtree)} for all $(p,q) \in M$,
    \begin{itemize}
        \item \textcolor{MarkColor}{(paths only meet at start/end)} either there are paths $p' \in \mathcal{P}(T_1)$ and $q' \in \mathcal{P}(T_2)$ such that $(p',q') \in M$ and $\pathstart(p) = \pathend(p')$ and $\pathstart(q) = \pathend(q')$, 
        \item or $\pathstart(p) = \troot(T_1)$ and $\pathstart(q) = \troot(T_2)$.
    \end{itemize}
\end{enumerate}
For an edge $e \in E(T_1)$ (or a vertex $v \in V(T_1)$), we write $e \notin M$ ($v \notin M$), if there is no pair $(p_1,p_2) \in M$ with $e \in p_1$ ($v \in p_1$). We use the same notation for edges or vertices of $T_2$.

For a path mapping $M$ between two abstract merge trees $T_1,\ell_1$ and $T_2,\ell_2$, we also define its corresponding edit operations $\mappingedit(M)$. They consist of the corresponding relabel, insert and delete operations:
\[
\begin{aligned}
\mappingrelabel(M) &= \{ (\ell_1(p_1),\ell_2(p_2)) \mid (p_1,p_2) \in M \}, \\
\mappinginsert(M) &= \{ (0,\ell_2(e_2)) \mid e_2 \in E(T_2),\ e_2 \notin M \}, \\
\mappingdelete(M) &= \{ (\ell_1(e_1),0) \mid e_1 \in E(T_1),\ e_1 \notin M \}.
\end{aligned}
\]
Then, we have $\mappingedit(M) = \mappingrelabel(M) \cup \mappinginsert(M) \cup \mappingdelete(M)$.
Moreover, we define the costs of a mapping via corresponding edit operations \textcolor{MarkColor}{and the path mapping distance as the costs of an optimal mapping}:
\begin{eqnarray}
\label{eq_pathmapping}
\cost(M) = \;\; \sum_{\mathclap{(l_1,l_2) \in \mappingedit(M)}} \;\; \cost(l_1,l_2), \;\;
\textcolor{MarkColor}{\pmdist(T_1,T_2) = \min_{M \text{ between } T_1,T_2} \{c(M)\}.}
\end{eqnarray}

\end{definition}

\input{wassersteinDistance}

\section{Deformation Based Geodesics and Barycenters}
\label{sec:method}

In this section, we describe the construction of geodesics and barycenters based on the path mapping distance.
We first give an intuition for
the barycenter construction before describing the algorithms in detail.
While the barycenter and geodesic construction is more technical than for the Wasserstein barycenters, the intuition behind it is similar and should be easy to grasp.
Furthermore, our algorithm does not depend on a prior normalization to obtain valid merge trees after interpolation.
Note that the geodesic is just a special case of the barycenter (as it is in the Wasserstein barycenter framework).
Thus, we first focus on the more generic case: we give an intuitive and algorithmic description of the barycenter construction.
In \autoref{sec:geodesic_app} of the suppl.\ material, we then formalize the interpolation which results for the case of two input trees (i.e.\ the resulting path of merge trees) and show that it indeed forms a geodesic in the metric space based on the path mapping distance.

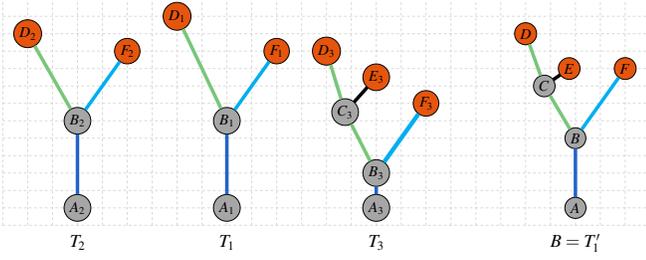
\begin{figure}[t!]
  \centering
   \resizebox{0.98\linewidth}{!}{
  \begin{tikzpicture}[yscale=0.7]
  
  \draw[help lines, color=gray!30, dashed] (-3,-1) grid (23,12);
  
  \definecolor{customred}{RGB}{230,85,13}
  \definecolor{customblue}{RGB}{33,100,200}
  \definecolor{customgreen}{RGB}{120,198,121}
  
  
  \node[draw,circle,fill=gray!70] at (6+0, 0) (root) {\textbf{\huge $A_1$}};
  \node[draw,circle,fill=gray!70] at (6+0, 5) (s1) {\textbf{\huge $B_1$}};
  
  \node[draw,circle,fill=customred] at (6-2, 11) (m1) {\textbf{\huge $D_1$}};
  \node[draw,circle,fill=customred] at (6+2, 9) (m2) {\textbf{\huge $F_1$}};
  
  \draw[customblue,line width=4pt] (root) -- (s1);
  \draw[customgreen,line width=4pt] (s1) -- (m1);
  \draw[cyan,line width=4pt] (s1) -- (m2);
  
  \node[font=\Huge] at (6+0,-2) (l) {$T_1$};
  
  
  \node[draw,circle,fill=gray!70] at (0+0, 0) (root') {\textbf{\huge $A_2$}};
  \node[draw,circle,fill=gray!70] at (0+0, 5) (s1') {\textbf{\huge $B_2$}};
  
  \node[draw,circle,fill=customred] at (0-2, 10) (m1') {\textbf{\huge $D_2$}};
  \node[draw,circle,fill=customred] at (0+2, 9) (m2') {\textbf{\huge $F_2$}};
  
  \draw[customblue,line width=4pt] (root') -- (s1');
  \draw[customgreen,line width=4pt] (s1') -- (m1');
  \draw[cyan,line width=4pt] (s1') -- (m2');
  
  \node[font=\Huge] at (0+0,-2) (l') {$T_2$}; 
  
  
  \node[draw,circle,fill=gray!70] at (12+0, 0) (root'') {\textbf{\huge $A_3$}};
  \node[draw,circle,fill=gray!70] at (12+0, 2) (s1'') {\textbf{\huge $B_3$}};
  \node[draw,circle,fill=gray!70] at (12-1.25, 5.5) (s2'') {\textbf{\huge $C_3$}};
  
  \node[draw,circle,fill=customred] at (12-2, 9) (m1'') {\textbf{\huge $D_3$}};
  \node[draw,circle,fill=customred] at (12+0, 7.5) (m2'') {\textbf{\huge $E_3$}};
  \node[draw,circle,fill=customred] at (12+2, 6) (m3'') {\textbf{\huge $F_3$}};
  
  \draw[customblue,line width=4pt] (root'') -- (s1'');
  \draw[customgreen,line width=4pt] (s1'') -- (s2'');
  \draw[customgreen,line width=4pt] (s2'') -- (m1'');
  \draw[black,line width=4pt] (s2'') -- (m2'');
  \draw[cyan,line width=5pt] (s1'') -- (m3'');
  
  \node[font=\Huge] at (12+0,-2) (l'') {$T_3$};
  
  
  \node[draw,circle,fill=gray!70] at (20+0, 0) (root''') {\textbf{\huge $A$}};
  \node[draw,circle,fill=gray!70] at (20+0, 4) (s1''') {\textbf{\huge $B$}};
  \node[draw,circle,fill=gray!70] at (20-1.25, 7) (s2''') {\textbf{\huge $C$}};
  
  \node[draw,circle,fill=customred] at (20-2, 10) (m1''') {\textbf{\huge $D$}};
  \node[draw,circle,fill=customred] at (20-0.25, 8) (m2''') {\textbf{\huge $E$}};
  \node[draw,circle,fill=customred] at (20+2, 8) (m3''') {\textbf{\huge $F$}};
  
  \draw[customblue,line width=4pt] (root''') -- (s1''');
  \draw[customgreen,line width=4pt] (s1''') -- (s2''');
  \draw[cyan,line width=4pt] (s1''') -- (m3''');
  \draw[customgreen,line width=4pt] (s2''') -- (m1''');
  \draw[black,line width=4pt] (s2''') -- (m2''');
  
  \node[font=\Huge] at (20+0,-2) (l''') {$B=T_1'$};
  
  \end{tikzpicture}
  }
  \caption{Example of barycenter assignment and update for member trees $T_1,T_2,T_3$ and initial candidate $T_1$. Optimal path mappings are illustrated through the edge colors. Edge lengths can be read from the grid.}
  \label{fig:intuition_barycenter_construction1}
\end{figure}

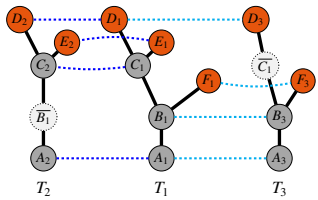
\begin{figure}[b!]
  \centering
   \resizebox{\linewidth}{!}{
  \begin{tikzpicture}[yscale=0.65]
  
  \draw[help lines, color=gray!0, dashed] (-9,1) grid (19,9);
  
  \definecolor{customred}{RGB}{230,85,13}
  \definecolor{customblue}{RGB}{33,100,200}
  \definecolor{customgreen}{RGB}{120,198,121}
  
  
  \node[draw,circle,fill=gray!70] at (5+0, 0) (root) {\textbf{\huge $A_1$}};
  \node[draw,circle,fill=gray!70] at (5+0, 3-0.3) (s1) {\textbf{\huge $B_1$}};
  \node[draw,circle,fill=gray!70] at (5-1, 7-0.3-0.65) (s2) {\textbf{\huge $C_1$}};
  
  \node[draw,circle,fill=customred] at (5-2, 10-0.3-0.65) (m1) {\textbf{\huge $D_1$}};
  \node[draw,circle,fill=customred] at (5+0, 8.5-0.3-0.65) (m2) {\textbf{\huge $E_1$}};
  \node[draw,circle,fill=customred] at (5+2, 6-0.3-0.65) (m3) {\textbf{\huge $F_1$}};
  
  \draw[black,line width=4pt] (root) -- (s1);
  \draw[black,line width=4pt] (s1) -- (s2);
  \draw[black,line width=4pt] (s2) -- (m1);
  \draw[black,line width=4pt] (s2) -- (m2);
  \draw[black,line width=4pt] (s1) -- (m3);
  
  \node[font=\Huge] at (5+0,-2) (l) {$T_1$};
  
  
  \node[draw,circle,fill=gray!70] at (0+0, 0) (root') {\textbf{\huge $A_2$}};
  \node[draw,dotted, very thick,circle,fill=gray!10] at (0, 3-0.3) (s1') {\textbf{\huge $\overline{B_1}$}};
  \node[draw,circle,fill=gray!70] at (0+0, 7-0.3-0.65) (s2') {\textbf{\huge $C_2$}};
  
  \node[draw,circle,fill=customred] at (0-1, 10-0.3-0.65) (m1') {\textbf{\huge $D_2$}};
  \node[draw,circle,fill=customred] at (0+1, 8.5-0.3-0.65) (m2') {\textbf{\huge $E_2$}};
  
  \draw[black,line width=4pt] (root') -- (s1');
  \draw[black,line width=4pt] (s1') -- (s2');
  \draw[black,line width=4pt] (s2') -- (m1');
  \draw[black,line width=4pt] (s2') -- (m2');
  
  \node[font=\Huge] at (0+0,-2) (l') {$T_2$};
  
  
  \node[draw,circle,fill=gray!70] at (10+0, 0) (root'') {\textbf{\huge $A_3$}};
  \node[draw,circle,fill=gray!70] at (10+0, 3-0.3) (s1'') {\textbf{\huge $B_3$}};
  \node[draw,dotted, very thick,circle,fill=gray!10] at (10-0.6, 7-0.3-0.65) (s2'') {\textbf{\huge $\overline{C_1}$}};
  
  \node[draw,circle,fill=customred] at (10-1, 10-0.3-0.65) (m1'') {\textbf{\huge $D_3$}};
  \node[draw,circle,fill=customred] at (10+1, 6-0.3-0.65) (m2'') {\textbf{\huge $F_3$}};
  
  \draw[black,line width=4pt] (root'') -- (s1'');
  \draw[black,line width=4pt] (s1'') -- (s2'');
  \draw[black,line width=4pt] (s2'') -- (m1'');
  \draw[black,line width=4pt] (s1'') -- (m2'');
  
  \node[font=\Huge] at (10+0,-2) (l'') {$T_3$};
  
  
  \draw[dashed,line width=2.5pt,blue] (root) to[bend left=0] (root');
  \draw[dashed,line width=2.5pt,blue] (s2) to[bend left=10] (s2');
  \draw[dashed,line width=2.5pt,blue] (m1) to[bend left=0] (m1');
  \draw[dashed,line width=2.5pt,blue] (m2) to[bend right=15] (m2');
  
  \draw[dashed,line width=2.5pt,cyan] (root) to[bend left=0] (root'');
  \draw[dashed,line width=2.5pt,cyan] (s1) to[bend left=0] (s1'');
  \draw[dashed,line width=2.5pt,cyan] (m1) to[bend left=0] (m1'');
  \draw[dashed,line width=2.5pt,cyan] (m3) to[bend right=15] (m2'');
  
  \end{tikzpicture}
  }
  \caption{Example of the barycenter update: Three merge trees $T_1$, $T_2$ and
  $T_3$ and two path mappings (dashed lines) are shown. $T_2$ and $T_3$
  contain an imaginary node, highlighted through the dotted
  strokes.}
  \label{fig:intuition_barycenter_construction2}
\end{figure}

\textbf{Intuition.}
The core idea of our approach is identical to the Wasserstein barycenters: \textcolor{MarkColor}{after initializing with a random member,} we alternate an \emph{assignment} and \emph{update} step and continue the iteration until a stable barycenter is reached. 
Given a barycenter candidate and the ensemble of merge trees, the assignment step computes the optimal path mapping between the candidate and each member tree.
The update step interpolates the lengths of mapped paths, where unmapped paths \textcolor{MarkColor}{are interpolated with imaginary edges of length~$0$.}
Thus, our approach is an adaptation of the Wasserstein barycenter method where we replace interpolation of mapped branches by interpolation of mapped paths.

An example can be seen in \autoref{fig:intuition_barycenter_construction1}.
We interpolate the paths $A_1B_1$ (length $5$), $A_2B_2$ (length $5$) and $A_3B_3$ (length $2$) to $AB$ (length $4$).
We interpolate the paths $B_1D_1$ (length $5$), $B_2D_2$ (length $6$) and $B_3C_3D_3$ (length $7$) to $BCD$ (length $6$).
The node $C$ in the barycenter is kept in the same relative position as $C_3$ in $T_3$.
The intuition behind this, in terms of the corresponding deformation, is that shortening $B_3C_3D_3$ to $BCD$ can also be seen as uniformly shortening both $B_3C_3$ to $BC$ and $C_3D_3$ to $CD$, which keeps the relative position constant.
The path $BF$ is handled exactly as $AB$.
As a last step, we shorten $C_3E_3$ to $\frac{1}{3}$ of its length, which intuitively means interpolating with two \textcolor{MarkColor}{imaginary edges of length $0$ (as it does not appear in the other two trees).}

However, this manner of interpolation is not necessarily well-defined for all combinations of path mappings.
The partition of the edges of the barycenter candidate may vary between the different mappings, which might make a meaningful interpolation impossible.
An example is shown in \autoref{fig:intuition_barycenter_construction2}.
Here, the setup is similar to \autoref{fig:intuition_barycenter_construction1}.
The mappings (here illustrated through the induced vertex maps) between $T_1$ and $T_2$ ($M_2$) and between $T_1$ and $T_3$ ($M_3$) are considered.
$M_2$ contains the path $A_1B_1C_1$, while $M_3$ only contains $B_1C_1$.
Here, it is unclear which of the paths $A_1B_1$ or $B_1C_1$ should be interpolated.

We fix this problem by splitting the mapped paths through insertion of imaginary nodes into the member trees until a proper interpolation is possible, i.e.\ until all mappings are pure edge mappings.
These splits can be designed such that the resulting fine-grained edge-mappings are still structure-preserving and thus can be interpolated into a meaningful merge tree.
\autoref{fig:intuition_barycenter_construction2} shows these imaginary nodes in the example trees, too.
This yields the following adapted mappings: $M_2$ now maps $A_1B_1$ to $A_2\overline{B_1}$ and $B_1C_1$ to $\overline{B_1}C_2$ instead of $A_1B_1C_1$, while $M_2$ now maps $B_1C_1$ to $B_3\overline{C_1}$ and $C_1D_1$ to $\overline{C_1}D_3$ instead of $B_1C_1D_1$.

Note that the example only considers the case where the two contradicting paths follow the same ``direction''.
They could also branch away from each other, e.g.\ one mapping could contain $A_1B_1C_1$ whereas the other could contain $A_1B_1F_1$.
Indeed, this case is also solved by splitting the paths until a pure edge mapping is reached.

In the following, we describe the barycenter algorithm in detail.
In contrast to the Wasserstein barycenters, our algorithm does not necessarily guarantee monotonic decrease of the Fréchet energy between its iterations.
However, in \autoref{sec:convergence}, we show empirical convergence \textcolor{MarkColor}{to a local energy minimum} on example datasets.

\textbf{Barycenters Construction.}
Given a barycenter candidate $B$ and path mappings $M_i \subseteq \mathcal{P}(B) \times \mathcal{P}(T_i)$ for each $1 \leq i \leq N$, we
build
the next candidate $B^*$ as follows.
First, any nodes or edges in $B$
not matched in any $M_i$ are removed.
Next, we relabel the nodes in $B$.
Lastly,
we insert all unmatched nodes and edges from the $M_i$ with adapted scalar values.
We describe these three procedures in more detail in
\textcolor{MarkColor}{\autoref{alg:all}.}

\textcolor{MarkColor}{The subroutine \texttt{RemoveUnmatched} in \autoref{alg:all}}
removes unmatched nodes from the barycenter candidate in
a straightforward manner by iterating over all mappings to check which edges
are part of mapped paths and removing those that do not appear. 

\textcolor{MarkColor}{The subroutine \texttt{RelabelBarycenter} in \autoref{alg:all}}
is more involved.
Recall that we want to split the mapped paths in the mappings through imaginary nodes to obtain pure edge mappings.
The algorithm does this implicitly.
It again iterates over all mappings and all pairs of paths in these mappings.
Let $(p,p')$ be such a pair where $p$ is a path in the barycenter candidate and $p'$ is a path in a member tree.
For each barycenter edge $e$ in $p$, it computes the length of the corresponding segment in $p'$, i.e.\ the same fraction of length.
In particular, the segment that corresponds to $e$ in $p'$ has length $\frac{\ell(e)}{\ell(p)} \dot \ell(p')$.
Then, the algorithm computes the average of all these segments.

As a last step,
\textcolor{MarkColor}{subroutine \texttt{AddUnmatched} in \autoref{alg:all}}
adds unmatched subtrees of the member trees to the barycenter.
Note that these subtrees are always complete subtrees rooted in some edge, since the path mapping distance only allows for insertions and deletions of leaf edges.
The algorithm also iterates over all path mapping and all pairs of paths in them.
For a mapped pair $(p,p')$, $p$ being the path in the barycenter, we add each node $v \in p'$ to $p$ and define its scalar value such that its relative position to start and end point is the same in $p$ and $p'$.
Each child of $v$ that is not on $p'$ is the root of a deleted subtree.
Thus, we insert these trees and define the length (in the barycenter) of each inserted edge to be $\frac{1}{k}$ of its original length (in the member tree).

As for the Wasserstein barycenters in~\cite{pont_vis21}, the iteration resembles Lloyd's method and is a generalization of the geodesic.
In contrast to the iteration for Wasserstein barycenters, this update procedure does not necessarily decrease the Fréchet energy \textcolor{MarkColor}{under the metric $\pmdist$ (defined as $ \sum_{i = 1}^{N} \pmdist(B,T_i)$)} in each step.
Thus, we opted to show the convergence \textcolor{MarkColor}{(to a local energy minimum)} experimentally, see \autoref{sec:convergence}.

To see why the update procedure does not necessarily decrease the Fréchet energy, consider Eqs.~\ref{eq_pathmapping} and~\ref{eq_wasserstein}.
Note that to obtain the path mapping distance, we simply take the sum of the single edit costs, whereas for the Wasserstein distance they are first squared and then we take the square root after summing up.
Since the mean value minimizes squared distances, not the sum of distances, the
update procedure given in
\autoref{alg:all}
is locally (i.e.,\ for a set of interpolated edges) not optimal.

The sum of distances is minimized by the median, not the mean.
Thus, we can take the median instead in
\texttt{RelabelBarycenter} and \texttt{AddUnmatched}
to circumvent this problem.
The resulting barycenter is, however, of inferior quality than the one based on the mean.
Using the median implies skipping
\texttt{AddUnmatched}
completely (the median will always be one of the $k-1$ imaginary edges of length~$0$), which means no edges are ever added to the barycenter.
If the initial candidate does not contain all features present in the ensemble, we get a poor quality barycenter, which we showcase on an example in \autoref{sec:heated_cylinder}.

\SetKwComment{Comment}{/* }{ */}
\setlength{\algomargin}{5pt}
\SetAlgoVlined
\SetInd{0.3em}{0.6em}

\begin{algorithm}[!t]
\caption{Subroutines of the update phase}
\label{alg:all}
\color{MarkColor}
\SetKwFunction{removeUnmatched}{RemoveUnmatched}
\SetKwProg{Fn}{Function}{:}{}
\DontPrintSemicolon
\Fn{\removeUnmatched($B,M_1,\dots,M_k,T_1,\dots,T_k$)}{
$B' = B$\;
Initialize EdgesMapped with $0$ for each edge in $E(B')$\;
\ForEach{$i=1 \dots k$}{
  \ForEach{$(p,p') \in M_i$}{
    \ForEach{\textup{Edge} $e \in p$}{
      EdgesMapped[$e$] = 1\;
    }
  }
}
\ForEach{\textup{Edge} $e =(c,p) \in E(B')$ \textup{in post-order}}{
  \If{\textup{EdgesMapped[$e$] = 0}}{
    Remove $e$ from $E(B')$ and $c$ from $V(B')$\;
  }
}
\ForEach{\textup{Node} $v \in V(B')$}{
  \If{$\deg(v)=1$}{
    Replace $(c,v) \in E(B')$ with $(c,p)$ for $p$ the parent of $v$\;
  }
}
\Return $B'$\;
}

\SetKwFunction{relabelBarycenter}{RelabelBarycenter}
\SetKwProg{Fn}{Function}{:}{}
\DontPrintSemicolon
\Fn{\relabelBarycenter($B,M_1,\dots,M_k,T_1,\dots,T_k$)}{
$B' = B$\;
Initialize EdgeLengths as $0$ for each edge in $E(B')$\;
\ForEach{$i=1 \dots k$}{
  \ForEach{$(p,p') \in M_i$}{
    \ForEach{\textup{Edge} $e \in p$}{
      EdgeLengths[$e$] += $\frac{\ell(e)}{\ell(p)} \cdot \ell(p')$ \;
    }
  }
}
\ForEach{\textup{Edge} $e \in E(B')$}{
  Set $\ell(e)$ to EdgeLengths[e] $\cdot \frac{1}{k}$ \;
}
\Return $B'$\;
}

\SetKwFunction{addUnmatched}{AddUnmatched}
\SetKwProg{Fn}{Function}{:}{}
\DontPrintSemicolon
\Fn{\addUnmatched($B,M_1,\dots,M_k,T_1,\dots,T_k$)}{
$B' = B$\;
\ForEach{$i=1 \dots k$}{
  \ForEach{$(p,p') \in M_i$}{
    \ForEach{\textup{Node} $v \in p'$}{
      Add $v$ to $p$ in $B'$ with $f_\ell(v)$ such that $\frac{f_\ell(v)}{\ell(p)} = \frac{f_\ell(v)}{\ell(p')}$ \;
      \ForEach{$e=(c,v) \in E(T_i)$}{
        \If{$c \notin p$}{
          Add the tree $T_e$ rooted in $e$ to $B'$ with labels $\frac{1}{k}\ell(e')$ for each edge $e' \in E(T_e)$ \;
        }
      }
    }
  }
}
\Return $B'$\;
}
\end{algorithm}

\begin{figure*}[t!]
  \centering
  \includegraphics[angle=270,width=150mm]{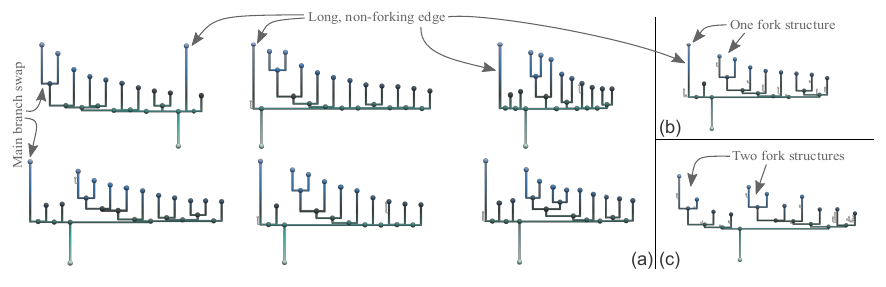}
  \caption{The six member split trees of the starting vortex ensemble with the two barycenters on the right. Branches of low persistence are uncolored and drawn thinner. All member trees (a) and the path mapping barycenter (b) contain one edge of high persistence without a fork structure as well as a fork structure of slightly lower persistence. The long edge also forms the main branch of the branch decomposition in all but one of the member trees. In contrast to that, the Wasserstein barycenter (c) creates a fork structure within this main branch, thus having two high-persistence forks. The reason is that the fork structure is the main branch in one member which leads to bad mappings.}
  \label{fig:barycenter_startingvortex}
\end{figure*}

\begin{figure}[b!]
  \centering
  \includegraphics[angle=270,width=84mm]{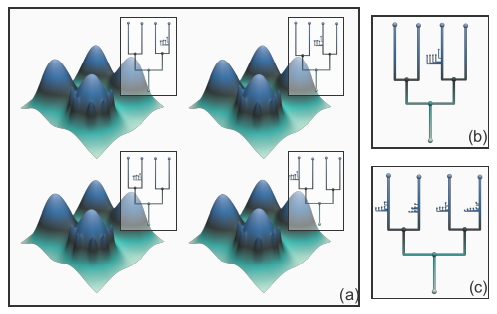}
  \caption{Four member fields from the analytical example with their split trees, shown in (a). On the right, the two barycenters are shown. The barycenter in (b) was computed using the path mapping distance, the one in (c) using the Wasserstein distance.}
  \label{fig:barycenter_falseClusters}
\end{figure}

\section{Experiments}
\label{sec:experiments}

In this section, we study the utility of our methods on three visualization and
analysis tasks: ensemble summarization, ensemble clustering, and time series temporal
reduction. For each, we compare results obtained using the path mapping distance to results achieved with the Wasserstein distance.
We briefly show the advantages of both barycenter methods over contour tree
alignments in the appendix. 

We describe implementation and experiment setup for each task in detail, and
study five datasets as well as convergence and performance.

\subsection{Implementation and Experiment Setup}

\begin{figure*}[]
\centering
\includegraphics[width=170mm]{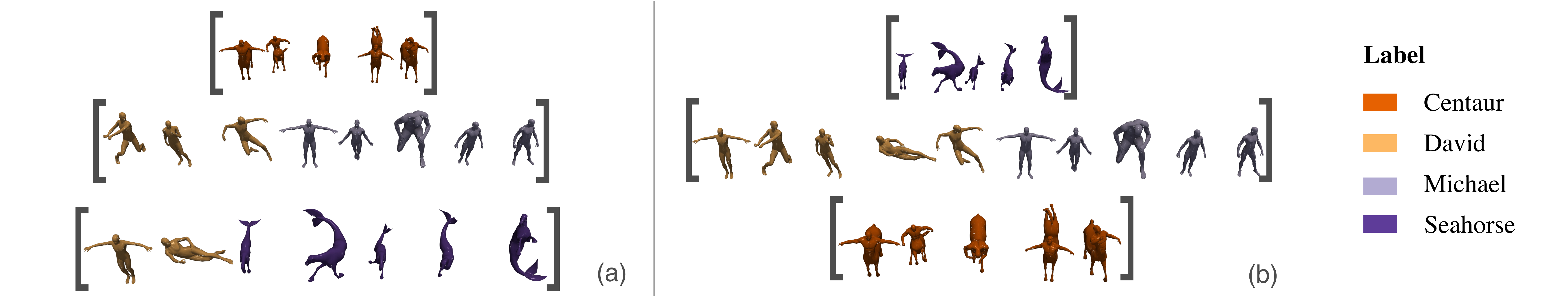}
\caption{Two screenshots of the clustering output for the TOSCA ensemble in ParaView. The meshes are colored by their shape name (note that the human cluster consists of two different shapes) and arranged according to the assigned cluster. The screenshot in (b) shows a correct result, which assigns all human shapes into one cluster, as well as all centaurs and all seahorses. It was computed using the path mapping barycenters. The screenhot in (a) shows an incorrect clustering computed by the Wasserstein barycenters.}
\label{fig:tosca_clustering}
\end{figure*}

Since the geodesic and barycenter constructions described in the previous section are just an adaptation of the Wasserstein geodesic and barycenter algorithms, we integrated our implementation into the original one in TTK~\cite{DBLP:journals/tvcg/TiernyFLGM18}.
In both cases, the barycenter computation is a generalization of the geodesic computation.
Thus, it suffices to adapt the barycenter method.
We extended the existing barycenter filter in TTK by adding our algorithm as an alternative option.

This enables us to use them as a drop-in in more advanced methods based on geodesics and barycenters as well.
TTK contains filters to compute a k-means~\cite{elkan03,celebi13} clustering based on merge tree distances and barycenters as well as filters for temporal reduction of merge tree sequences and the corresponding reconstruction.
Details on these methods can be found in~\cite{pont_vis21}.
In all of them, it is now possible to use the path mapping distance instead of the Wasserstein distance, which enables users to reuse old workflows when applying the new method.
For our experiments, we used the TTK implementation and ParaView~\cite{paraviewBook} on the following tasks (executed on an Intel Core i7-7700 with 64GB of RAM).

\textbf{Ensemble Summarization.} Given an ensemble of $k$ scalar fields $f_1,f_2,\dots,f_k$, we first compute the $k$ merge trees $T_1,\dots,T_k$.
Then, we compute the barycenter merge tree $B$ of $T_1,\dots,T_k$.
This barycenter tree should visually summarize the ensemble to give the user an overview of the existing features and the overall topological structure of the scalar fields in it.
The barycenter tree should contain all important features of the members and should not contain prominent structures that cannot be mapped back to features in at least some of the members.
We will rate the results on a purely visual basis.

\textbf{Ensemble Clustering.} Using the iterative barycenter computation, it is possible to apply existing clustering strategies such as the k-means algorithm.
We added our implementation of the path mapping distance to the existing merge tree clustering framework in
TTK~\cite{pont_vis21}.

Then, given an ensemble of $k$ scalar fields $f_1,f_2,\dots,f_k$, we first compute the $k$ merge trees $T_1,\dots,T_k$.
Next, we apply the k-means algorithm to $T_1,\dots,T_k$.
We compare the resulting cluster assignment to a previously known ground truth.
We consider the result as correct, if no two members assigned to the same cluster belong to different ground truth clusters.
\textcolor{MarkColor}{An addition to counting the number of fully correct results, we compute the adjusted rand index~\cite{hubert1985comparing} (ARI).}
Since the k-means implementation in TTK can utilize randomized initialization, we did multiple runs of both the path mapping based and the Wasserstein distance based clustering.
\textcolor{MarkColor}{We compare the percentage of correct runs as well as the average ARI.}

\textbf{Temporal reduction.} Given an ensemble of $k$ scalar fields $f_1,f_2,\dots,f_k$, we first compute the $k$ merge trees $T_1,\dots,T_k$.
This time series of merge trees is then reduced to $k'$ key frames.
The target number $k'$ is given as an input by the user.
We greedily remove merge trees from the time series, until only $k'$ merge trees are left, as described in~\cite{pont_vis21}.
To reconstruct the original time series, we compute geodesic merge trees of the adjacent key frames.

To rate the quality of the reduction and reconstruction, we compute the distance between the original merge trees and the corresponding reconstructed ones.
We compute the key frames and the reconstruction with both the path mapping distance and the Wasserstein distance and compare quality of the two reconstructions.
The comparison can be done visually (i.e.\ do the reconstructed merge trees look like good interpolations) or based on the distance between the original and reconstructed trees.
The latter is done using both distance measures.

\smallskip
In the remainder of this section, we study the behavior of the path mapping based approaches on these three tasks.
We go through five different datasets and discuss the results in the corresponding subsection.
As a last step, we study convergence and running times in practice.

\subsection{Analytical Example}

The analytical example is an ensemble consisting of 20 members.
Each member is a 2-dimensional regular grid with a scalar function defined on the vertices.
It was first used in~\cite{DBLP:journals/cgf/WetzelsLG22} to showcase the advantages of branch decomposition-independent edit distances over those using fixed BDTs.
It is a synthetic dataset designed specifically to provoke instabilities for BDT-based methods, which we explain in the following.

Each member field consists of four main peaks, one of which has five smaller peaks arising from it (see \autoref{fig:barycenter_falseClusters}, together with the corresponding merge trees).
The positions of the maxima as well as their heights are all chosen randomly within small ranges, leading to differences in the nesting of the persistence-based branch hierarchy.
More specifically, the order of the four main maxima is ``shuffled'' in the BDT.
In contrast, the five side maxima (although also slightly perturbated) stay on the same hill. 
As a consequence, the corresponding noisy sub-tree (located on the front-most hill in \autoref{fig:barycenter_falseClusters}) travels in the BDTs, depending on the maximum value reached by its main hill.
Therefore, structure-preserving mapping on BDTs can either map all four main peaks correctly or the five side peaks, but not both.

In \autoref{fig:barycenter_falseClusters}, we illustrate the difference in the BDTs through the position of the corresponding arcs in the planar layout.
They are placed according to the order in the branch decomposition from left to right.
In the remainder of this paper, we will refer to this behavior as a maximum swap.
Note that each of the 20 member trees is of the form of one of the four merge trees in \autoref{fig:barycenter_falseClusters} and we omit to show them all.

We used this dataset for the summarization task, applying both the path mapping and the Wasserstein distance.
The resulting path mapping barycenter strongly resembles the member trees: it has four main maxima and there is exactly one of them which has side branches leading to smaller maxima.
The barycenter based on the Wasserstein distance duplicates the side branches: each main maximum has attached some side peaks.
The reason is that the Wasserstein distance utilizes structure-preserving mappings on BDTs, which fail as explained above.
In contrast, the path mapping distance is branch decomposition-independent, working purely on the merge trees (which are highly similar).
Thus, the Wasserstein barycenter is a less precise summarization of the orginal member trees.
\autoref{fig:barycenter_falseClusters} also shows the two barycenters.

\begin{figure}[b!]
  \centering
  \includegraphics[width=76mm]{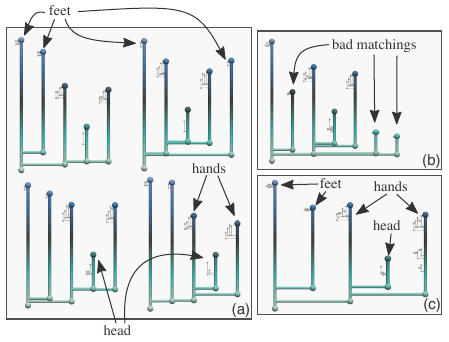}
  \caption{Dataset summarization on the TOSCA dataset: four example members of the human cluster are shown in (a), the path mapping barycenter of this cluster can be seen in (b), the Wasserstein barycenter in (c). The Wasserstein barycenter is a good representative of the ensemble with clearly identifiable head, hands and feet. The path mapping fails to do so and contains branches that are hard to interpret, stemming from poor quality mappings.}
  \label{fig:barycenter_tosca}
\end{figure}

\subsection{Starting Vortex}

The starting vortex dataset is an ensemble of 2D regular grids with a scalar function on the vertices representing flow turbulence behind a wing.
It was generated using the Gerris flow solver~\cite{gerrisflowsolver} and has been used in~\cite{favelier_vis18,vidal_vis19,pont_vis21}.
Each member represents a different inclination angle of the wing.
In~\cite{pont_vis21} it was successfully used for the clustering task, where the two clusters were defined by two different ranges of angles, consisting of six  consecutive integer values.
Here, we use one of these clusters and compute the barycenter split tree.
We applied a topological simplification with a threshold of 5\% of the scalar range.
\autoref{fig:barycenter_startingvortex} shows the member trees and both the path mapping barycenter and the Wasserstein barycenter.
The path mapping barycenter is clearly a better representative of the ensemble than the Wasserstein barycenter.
We should note that the behavior depends on the initial barycenter candidate.
For the path mapping distance, all six resulting barycenters are of good
quality, whereas for the Wasserstein distance, five of six initial candidates
produce low-quality results, again due to a maximum swap (cf.~\autoref{fig:barycenter_startingvortex}), as explained for the first dataset.
We provide additional renderings for all possible barycenter outcomes in \autoref{sec:startingvortex_app} of the suppl.\ material.

\begin{figure*}[t!]
  \centering
  \includegraphics[angle=270,width=150mm]{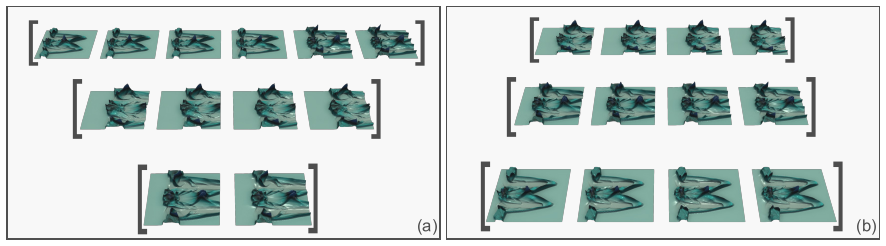}
  \caption{Two clustering results for the ionization front ensemble. The result for the path mapping distance is shown on the right. The assigned clusters correspond to the three ground truth phases of the simulation, which are visually clearly distinguishable. The earliest one is shown in the bottom cluster, the middle one in the center cluster, the last one in the top clusters. On the left, the clustering result for the Wasserstein distance can be seen. The mid phase is split up and mixed with the earliest phase (top). The last phase is identified correctly (center line).}
  \label{fig:clustering_ionization}
\end{figure*}

\subsection{TOSCA Shape Matching Ensemble}

The TOSCA non-rigid world dataset~\cite{DBLP:series/mcs/BronsteinBK09} is a shape matching ensemble containing several shapes (different humans and animals) in various poses.
The meshes have an average geodesic distance field~\cite{hilaga:sig:2001} attached.
We use the same preprocessed scalar fields as Sridharamurthy et al. in~\cite{SridharamurthyM20} and applied topological simplification with a threshold of 2\% of the scalar range in a pre-processing step.
We picked the first five poses for four different shapes (David, Michael, seahorse and centaur) and defined the following clusters as ground truth: the two human shapes David and Michael
form one cluster,
the seahorse another one,
as well as the centaur.

We applied the k-means clustering to this set of scalar fields and compared the assigned clusters to the ground truth.
We used a target number of 3 clusters for the basic experiment and performed runs for a target number of 4 as well.
In the latter case, we counted runs with split clusters as correct, but not if two clusters were mixed.

The path mapping distance clearly outperforms the Wasserstein distance.
\autoref{fig:tosca_clustering} shows two typical outputs in ParaView.
Since k-means implementation is randomized, we performed 100 runs for each algorithm and checked whether the assigned clusters are completely correct.
Here, the path mapping barycenters produced \textcolor{MarkColor}{57} correct clusterings \textcolor{MarkColor}{and an ARI of 0.75} when using a target number of 3, whereas the clustering based Wasserstein barycenters was never correct \textcolor{MarkColor}{and achieved an ARI of 0.34}.
Interestingly, we get an even better accuracy (in terms of completely correct runs) for a target number of 4.
Here, the number of correct runs for the path mapping distance was \textcolor{MarkColor}{78 and 1} for the Wasserstein distance.
\textcolor{MarkColor}{The ARIs were 0.7 and 0.43.}
Overall, we observed accuracies of the path mapping barycenters to be significantly better in comparison to the Wasserstein barycenters.

Furthermore, we took the ten human poses forming the largest cluster and computed the barycenter
tree as a representative.
Surprisingly, while path mappings show significantly better results for the clustering, they fail to produce a good representative for the human cluster.
\autoref{fig:barycenter_tosca} illustrates this behavior.
The reason is that there are saddle swap instabilities within the human shapes.
While both distances are unable to handle those \emph{in general} (although for both we can preprocess the data by merging close saddles), the Wasserstein distance is able to handle \emph{some} instabilities by working on unordered BDTs.
It thereby allows to match swapped saddles, as long as they are children of the same parent branch.
This is not possible for
path mappings.
In~\cite{DBLP:journals/cgf/WetzelsLG22}, this was discussed in more detail for
branch mappings
(but the arguments translate to path mappings as well).
There, it says that the search spaces of the branch mapping distance and Wasserstein distance are orthogonal to each other.
The subset of the TOSCA ensemble considered in this section is a good example for exactly this orthogonality.  

\subsection{Ionization Front}

The ionization front dataset~\cite{scivis2008} consists of 2D slices from a time dependent scalar field representing ion concentration.
The scalar fields were preprocessed with a topological simplification using a threshold of 5\% of the scalar range.
We used this dataset as a time series as well as an ensemble for clustering by picking three clusters (different phases of the simulation) of consecutive time points.

The examples presented here mainly focus around a specific main maximum swap between the time steps 126 and 127.
This main maximum swap is shown in \autoref{fig:teaser}.
There, it is easy to see that the Wasserstein distance fails to produce a good mapping, which also leads to a poor quality geodesic tree.
While the layouts of the three trees on the right are aligned to show the correct mapping, the layouts on the left are based on the branch decomposition.
The branch decomposition based layout shows the main maximum swap (in the first tree, the short fork is the most persistent branch feature, in the last one it is the long fork), which is the reason for the bad mapping.

In the following, we use this observation to showcase the influence on two
applications.
First, we show that the bad mapping leads to bad clustering
with the Wasserstein distance if the ground truth clusters contain the main maximum swap.
This is not the case for the path mapping distance.
Next, we show that the same holds for the temporal reduction: using the Wasserstein distance, more time steps are needed to get the same reconstruction quality as with the path mapping distance.

\begin{figure}[b!]
    \centering
    \includegraphics[width=0.62\linewidth]{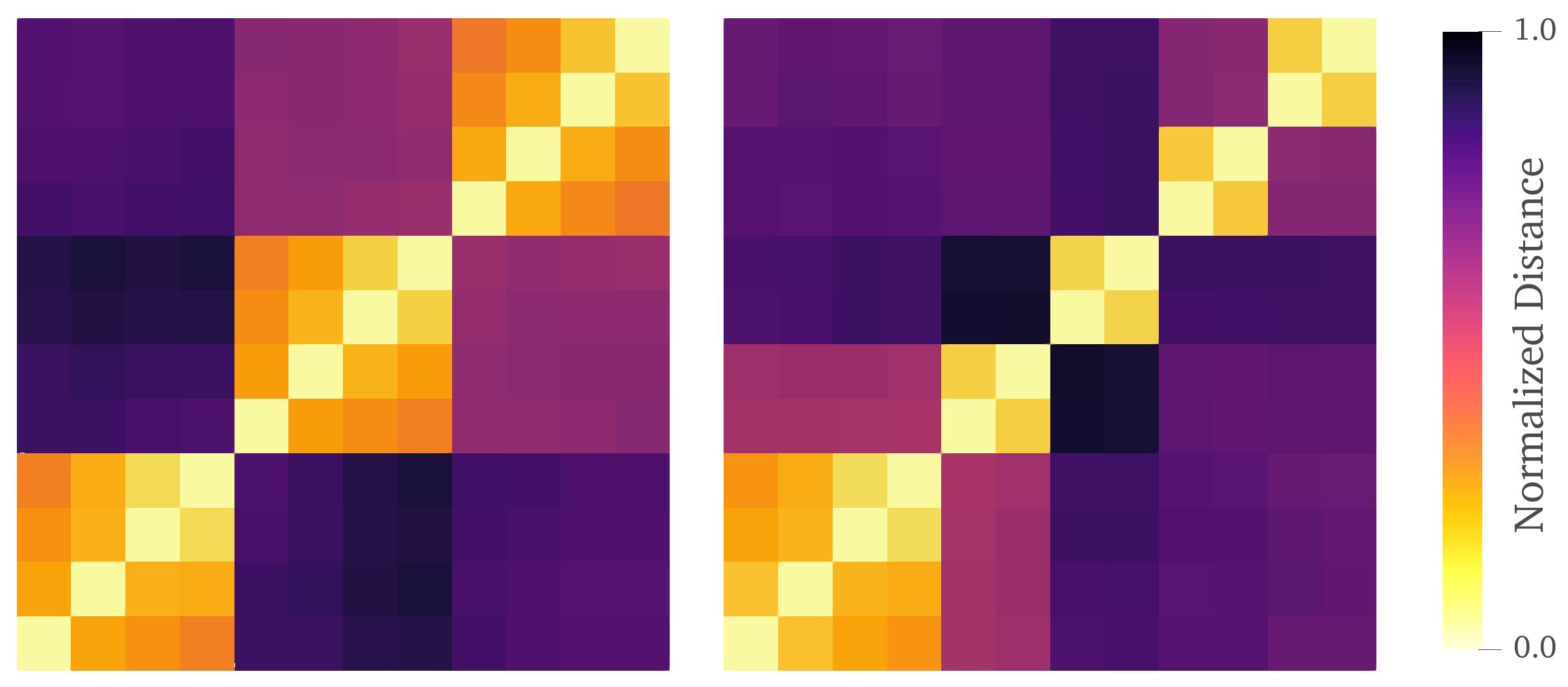}
    \caption{The distance matrices on the clustering ensemble for the path mapping distance (left) and the Wasserstein distance (right). The rows and columns are ordered by time step. The path mapping distance clearly shows the three ground truth clusters, whereas the Wasserstein distance has five clusters in total. This leads to the poor results when applying a k-means clustering.}
    \label{fig:distmatrix_ionization}
\end{figure}

For the clustering, we picked three sets of four consecutive time steps each.
We then applied the k-means algorithm using both the path mapping distance and the Wasserstein distance. 
Again, the path mapping distance shows significantly better results.
To better understand the reason for this, we also plotted the distance matrices for the two distances in \autoref{fig:distmatrix_ionization}.
It is easy to see that the Wasserstein distance splits two of the three ground truth clusters into two smaller clusters each, leading to five clusters in total.
\textcolor{MarkColor}{This happens due to a maximum swap within these clusters.
We observed 87\% of the runs with the path mapping distance to be correct, and 0\% with the Wasserstein distance.
The ARIs were 0.92 and 0.46.}

\begin{figure}[b!]
  \centering
  \includegraphics[angle=270,width=85mm]{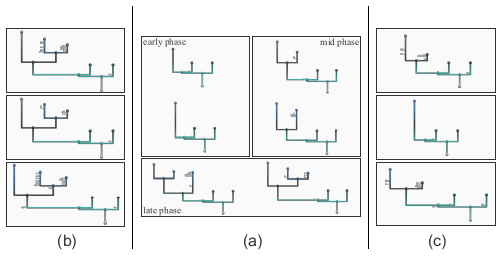}
  \caption{Six member trees from the heated cylinder ensemble (a) with several barycenters. The six example members come from three different phases of one single run. The barycenters on the right (c) were computed on a consecutive subsequence of a single run, whereas the barycenters on the left (b) were computed on one fixed time step from the late phase for each run. The top row barycenters in (b) and (c) were computed using the path mapping distance and the typical mean method. The center row barycenters are based on the path mapping distance in the median variant. The bottom row barycenters are the Wasserstein barycenters.}
  \label{fig:barycenter_heatedCylinder}
\end{figure}

\begin{figure*}[t!]
  \centering
  \includegraphics[width=170mm]{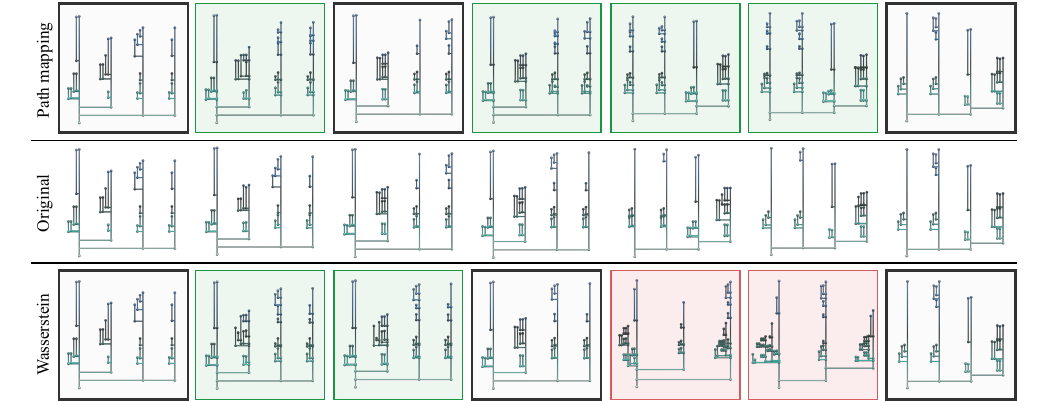}
  \caption{Two reconstructions (top and bottom line) of a time series of merge trees (center line) after temporal reduction. For both the Wasserstein distance and the path mapping distance, we used three key frames in the encoding phase (highlighted through black frames). We then used path mapping geodesics and Wasserstein geodesics to reconstruct the sequence. Visually good reconstructions are highlighted in green, bad ones in red. A numerical comparison can be found in \autoref{tab:distance} \textcolor{MarkColor}{of \autoref{sec:tempred_app} of the suppl.\ material}.}
  \label{fig:tempred_ionization}
\end{figure*}

For the temporal reduction task, \autoref{fig:tempred_ionization} illustrates the results of both methods on a subsequence around the described maximum swap.
The time series used here consists of 12 time steps around the above explained saddle swap: 6 time steps before and after the swap, respectively.
We used 3 as the target number of key frames.
As expected, in the first half of the series (before the swap), both methods yield very precise reconstructions, since in both cases the middle key frame is chosen \emph{before} the swap.
In contrast, in the second half, only the path mapping geodesic yields precise reconstructions, since the bad mapping of the Wasserstein distance (as discussed above) between the second and last keyframe also leads to bad interpolation.
In fact, the Wasserstein geodesic needs 4 key frames for a precise reconstruction (the first time step, the last, one right before, and one right after the swap), whereas the path mapping geodesic already gives good results for only 2 key frames (see \autoref{sec:tempred_app} of the suppl.\ material).

Apart from a visual or qualitative comparison, we also provide a quantitative comparison, i.e.\ we compared the actual distances to the original member trees.
To avoid a bias through the used distances, we provided both the path mapping and the Wasserstein distance for both reconstructions each.
\autoref{tab:distance} \textcolor{MarkColor}{of \autoref{sec:tempred_app} of the suppl.\ material} shows the results.
In the first half of the series, the distances are generally small.
In the second half, the path mapping geodesics yield much smaller distances, independent of the chosen distance measure.
\textcolor{MarkColor}{This intuitively fits with the observations in \autoref{fig:tempred_ionization}.}

\subsection{Heated Cylinder}
\label{sec:heated_cylinder}

The heated cylinder ensemble consists of 23 time-dependent scalar fields describing flow around a heated pole.
Each member was created using small-scale perturbations of the initial conditions.
It was used in~\cite{LohfinkWLWG20} for the computation of
fuzzy contour trees.
We again apply 
topological
simplification (5\% of the scalar range).
We compute the barycenter tree for both a fixed time point with varying runs \textcolor{MarkColor}{(23 trees)} and a fixed run with varying time points \textcolor{MarkColor}{(30 trees)}.
The results are shown in \autoref{fig:barycenter_heatedCylinder}.
The latter case shows that the median-based barycenter
fails if the initial
candidate contains few features, since missing features will never be added.
In contrast to that, on a fixed time step, all member trees are very similar and all three methods perform well.

We also compare the barycenter summarization of the ensemble to the fuzzy contour tree~\cite{LohfinkWLWG20} in \autoref{sec:heatedCylinder_app} of the suppl. material.
We used the publicly available notebook~\cite{fctCode} and the TTK implementation~\cite{LohfinkWLWG20,DBLP:journals/tvcg/TiernyFLGM18} of the contour tree alignment and fuzzy contour tree algorithms.

\subsection{Convergence and Runtime Performance}
\label{sec:convergence}

\begin{figure}[b!]
    \centering
    \includegraphics[width=0.9\linewidth]{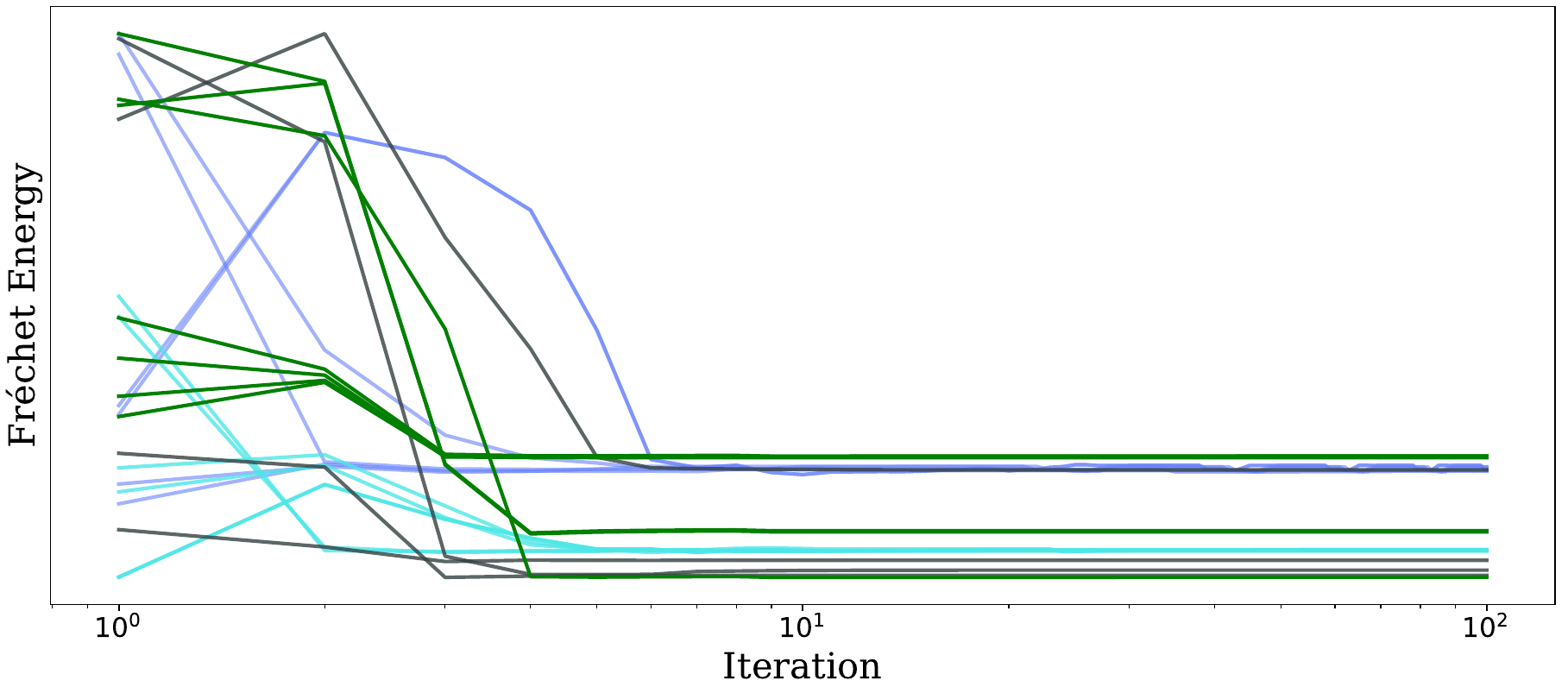}
    \caption{(Relative) Fréchet energy at each iteration of the barycenter algorithm for two
     member series of the heated cylinder ensemble (blue/cyan), the starting
     vortex ensemble (gray) and the human cluster of the TOSCA ensemble (green).
     Since absolute value of the energy depends on the given scalar function,
     we omit quantitative labels on the y-axis.}
    \label{fig:convergence}
\end{figure}

We now study the convergence of the barycenter iteration experimentally.
We ran the barycenter algorithm for 100 iterations on four different datasets and on each dataset we used different initial candidates.
The plots can be found in \autoref{fig:convergence}.
As can be seen, the Fréchet energy converges in less than 10 iterations on all datasets.
Although the first iteration often increases the energy, it is overall significantly decreased.
However, for some initial candidates of the heated cylinder ensemble, it converges at an energy higher than in the initial state.
Furthermore, the plots show that the convergence energy can differ within one dataset depending on the initial candidate.

\smallskip
Next, we consider the runtime performance/complexity of the new method.
Branch decomposition-independent edit distances have an asymptotic running time of $\mathcal{O}(n^4)$ and need the same amount of memory, in contrast to $\mathcal{O}(n^2)$ for classic tree edit distances.
This limits the practical application of these distances on very large merge trees containing thousands of nodes. E.g.\ distance computation on the unsimplified asteroid dataset~\cite{scivis2018} requires in excess of 500GB of RAM and should therefore be performed on advanced hardware.
Running times in the range of seconds for single distance computation are observed for merge trees of up to a few hundred nodes (see~\cite{DBLP:journals/cgf/WetzelsLG22,path_mappings}).
Since finding the geodesic only requires the computation of a single path mapping, this performance analysis can also be applied here.

For the barycenter computation, multiple path mappings have to be computed in each iteration.
The barycenter candidate (initially a member tree) can get significantly larger than the member trees, since it may contain many unmapped features.
Thus, we now focus on the runtime performance of the barycenter iteration, considering the amount of ensemble members, the size of the barycenter and the iteration number in addition to the size of the input trees.

In \autoref{tab:runtime}, running times for the barycenter computation is given for multiple ensembles.
Here, we used the datasets from above (with different levels of simplification) as well as a jet flow simulation dataset and the vortex street dataset from~\cite{SridharamurthyM20} (simulated by Weinkauf~\cite{weinkauf10c} using the Gerris flow solver~\cite{gerrisflowsolver}).
With the path mapping barycenters, feasible running times in the range of seconds are reached on all datasets with input trees of less than 100 nodes, even if the barycenter size gets up to over 400 nodes.
On the Jet dataset, with an average member tree size of roughly 150 and barycenter sizes of up to 400 vertices, the running times were still in the range of minutes.
\textcolor{MarkColor}{Regarding the effect of the ensemble size, a super-linear increase in runtime can be observed.
This is probably due to the fact that the barycenter size scales with the number of members: not only the number of distance computations increases, but their complexity does, too.
However, this happens for both the Wasserstein and the path mapping distance.}

The running times of the Wasserstein barycenter do not suffer from such an increase, as expected.
All running times are below 1s.
The comparison on the analytical example suggests that, in case of decreased barycenter sizes through the better mapping quality, the path mapping distance can sometimes be quicker.

Overall, the numbers suggest that a topological simplification should be applied prior to the path mapping barycenter computation.
Since one of the main purposes of the barycenter is a visual summarization, a few hundred vertices
seems to be a reasonable limitation, as merge trees containing more nodes are hard to use as a visual summary anyway.

\begin{table}[]
\caption{Runtime performance of the path mapping and Wasserstein barycenters for various datasets. $|T|$ denotes the average size of the member trees, $k$ the amount of members. The columns labeled $|B|$ show the minimum and maximum barycenter size, averaged over the different runs. The average number of iterations is denoted by $n$ and average runtime by $t$.}
\label{tab:runtime}
\centering
 \resizebox{\linewidth}{!}{
 \renewcommand{\arraystretch}{1.2}
\begin{tabular}{l||c|c||c|c|c||c|c|c}
Dataset & $|T|$ & $k$ & $|B_P|$ & $n_P$ & $t_P$ & $|B_W|$ & $n_W$ & $t_W$ \\
\hline
Analytic Ex. & 18 & 20 & 18-18 & 3 & 0.009s & 70-150 & 3 & 0.02s \\
Starting Vortex & 33 & 6 & 70-78 & 3.4 & 0.05s & 90-100 & 4.4 & 0.01s \\
TOSCA & 35 & 10 & 61-75 & 4.5 & 0.05s & 92-118 & 3.6 & 0.017s \\
TOSCA & 48 & 10 & 93-136 & 5 & 0.28s & 156-185 & 5 & 0.04s \\
Ionization & 47 & 12 & 105-196 & 6.3 & 0.77s & 167-261 & 5 & 0.07s \\
Ionization & 71 & 12 & 191-313 & 5.7 & 3.5s & 330-456 & 5 & 0.165s \\
Ionization & 89 & 12 & 257-406 & 4.8 & 14.5s & 448-579 & 4.3 & 0.22s \\
Vortex Street & 69 & 10 & 156-172 & 4.9 & 1.84s & 168-174 & 4.6 & 0.04s \\
Jet & 149 & 10 & 242-377 & 7 & 348s & 257-411 & 5.6 & 0.29s \\
\color{MarkColor}Heat. Cylinder & \color{MarkColor}19 & \color{MarkColor}23 & \color{MarkColor}76-145 & \color{MarkColor}5 & \color{MarkColor}0.24s & \color{MarkColor}129-174 & \color{MarkColor}3.8 & \color{MarkColor}0.03s \\
\color{MarkColor}Heat. Cylinder & \color{MarkColor}19 & \color{MarkColor}46 & \color{MarkColor}94-234 & \color{MarkColor}5.8 & \color{MarkColor}1.0s & \color{MarkColor}175-297 & \color{MarkColor}5.1 & \color{MarkColor}0.12s \\
\color{MarkColor}Heat. Cylinder & \color{MarkColor}19 & \color{MarkColor}69 & \color{MarkColor}125-405 & \color{MarkColor}5.6 & \color{MarkColor}3.7s & \color{MarkColor}304-477 & \color{MarkColor}3.8 & \color{MarkColor}0.24s 
\end{tabular}
}
\end{table}

\section{Results}
\label{sec:conclusion}

In this paper, we defined new notions of geodesics and barycenters of merge trees based on the metric space defined by the path mapping distance.
We gave an algorithm heuristically computing barycenters as well as accurate geodesics.
We implemented the algorithm in TTK and integrated it into the existing Wasserstein barycenter implementation, yielding a unified framework.
We then provided experimental evidence for the improved quality of path mapping barycenters over Wasserstein barycenters on five different datasets: the path mapping barycenters are often better summarizations of ensembles, lead more frequently to correct clustering results and can further reduce time series of merge trees while retaining or even improving reconstruction quality.
\textcolor{MarkColor}{Furthermore, we highlighted limitations that are summarized in the following paragraph. These should be considered in future work.}                 

\textcolor{MarkColor}{\textbf{Limitations.}}
\textcolor{MarkColor}{Our method shows increased runtimes, both asymptotically ($\mathcal{O}(n^4)$ vs $\mathcal{O}(n^2)$) and practically (single threaded times are in the range of minutes on merge trees with 100-200 nodes) when compared to the Wasserstein framework.}
\textcolor{MarkColor}{There is also the possibility to reduce result quality} on datasets containing specific forms of saddle swaps. 
\textcolor{MarkColor}{Furthermore, it remains open whether our barycenter algorithm fulfills some form of formal convergence property, or if this can be achieved with adapted strategies.
The influence of the random initialization should also be studied in more detail.}

\textcolor{MarkColor}{\textbf{Conclusion.}}
Overall, we gave strong evidence that path mapping barycenters (though having limitations) should be considered in practical applications and gave users the option to do so (e.g.\ choosing the preferred method based on sizes of the input trees) in an established framework.

\todo[inline]{Changes:\\
- added Limitations subsection in conclusions\\
- initialization for Wasserstein barycenters and path mapping barycenters\\
- added ARIs to experiments
- added intuitive description of geodesic and barycenter\\
- added formal definition of geodesic to supplement\\
- fixed several typos and minor mistakes\\
- added description of "imaginary edge" in intuition for barycenter computation\\
- minor change in teaser caption\\
- intuitive explanations for path mapping definition\\
- experiment for number of members\\
- merged algorithms to gain space\\
- moved Table 1 to appendix\\
- tons of minor changes regarding remarks by the reviewers\\
ToDos:\\
- Discuss effect of initial candidate? (I think we are doing this already, maybe note in cover letter)\\
- what about identical subtrees (R2) -> cover letter
}

\acknowledgments{
The authors wish to thank Raghavendra Sridharamurthy for providing the pre-processed TOSCA dataset and Marvin Petersen for helpful discussions. This work is funded by the Deutsche Forschungsgemeinschaft (DFG, German Research Foundation) – 442077441 - as well as by the European Commission grant ERC-2019-COG \emph{``TORI''} (ref. 863464, \url{https://erc-tori.github.io/}).}

\section*{Supplementary Material}

This manuscript is accompanied by supplementary material:

\begin{itemize}
    \item The publicly available source code~\cite{repository} is provided together with detailed instructions to compile it and reproduce the images shown in~\autoref{fig:teaser}. This implementation will be contributed as open source to TTK in the future.
    
    The companion ZIP file contains an archive of the repository.
    \item We provide a supplementary PDF that contains additional images and a proof that the proposed interpolation yields a geodesic.
\end{itemize}

\bibliographystyle{abbrv-doi-hyperref}

\bibliography{paper}

\makeatletter\@input{xxy.tex}\makeatother

\end{document}



\maketitle

\appendix

\section{Geodesic Proof}
\label{sec:geodesic_app}

In this section, we provide a formal proof that for two input merge trees the interpolation defined in \autoref{sec:method} indeed forms a geodesic between the input trees.
We now give a formal description of the interpolated geodesic trees to make formal arguments easier.
Let $(T_0,f_0),(T_1,f_1)$ be two merge trees, $M \subseteq \mathcal{P}(T_0) \times \mathcal{P}(T_1)$ a path mapping between $T_0$ and $T_1$ and let $(T_\alpha,f_\alpha)_{\alpha \in (0,1)}$ as follows.
We write $\ell_0,\ell_1,\ell_\alpha$ for $\ell_{f_0},\ell_{f_1},\ell_{f_\alpha}$.

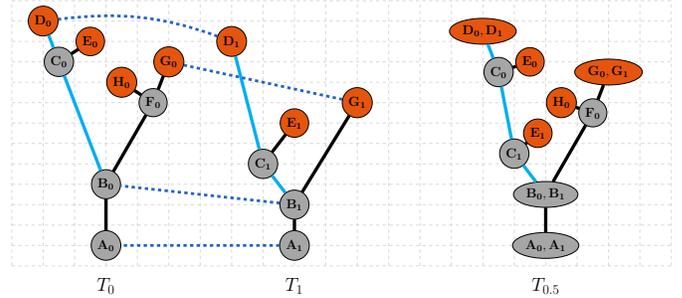
\begin{figure}
  \centering
   \resizebox{\linewidth}{!}{
  \begin{tikzpicture}[yscale=0.65]
  
  \draw[help lines, color=gray!30, dashed] (-3,-1) grid (18,12);
  
  \definecolor{customred}{RGB}{230,85,13}
  \definecolor{customblue}{RGB}{33,100,200}
  
  
  \node[draw,circle,fill=gray!70] at (6+0, 0) (root) {\Large $\mathbf{A_1}$};
  \node[draw,circle,fill=gray!70] at (6+0, 2) (s1) {\Large $\mathbf{B_1}$};
  \node[draw,circle,fill=gray!70] at (6-1, 4) (s2) {\Large $\mathbf{C_1}$};
  
  \node[draw,circle,fill=customred] at (6-2, 10) (m1) {\Large $\mathbf{D_1}$};
  \node[draw,circle,fill=customred] at (6-0, 6) (m2) {\Large $\mathbf{E_1}$};
  \node[draw,circle,fill=customred] at (6+2, 7) (m3) {\Large $\mathbf{G_1}$};
  
  \draw[black,line width=3pt] (root) -- (s1);
  \draw[cyan,line width=3pt] (s1) -- (s2);
  \draw[cyan,line width=3pt] (s2) -- (m1);
  \draw[black,line width=3pt] (s2) -- (m2);
  \draw[black,line width=3pt] (s1) -- (m3);
  
  \node[font=\huge] at (6+0,-2) (l) {$T_1$}; 
  
  
  \node[draw,circle,fill=gray!70] at (0+0, 0) (root') {\Large $\mathbf{A_0}$};
  \node[draw,circle,fill=gray!70] at (0+0, 3) (s1') {\Large $\mathbf{B_0}$};
  \node[draw,circle,fill=gray!70] at (0-1.5, 9) (s2') {\Large $\mathbf{C_0}$};
  \node[draw,circle,fill=gray!70] at (0+1.5, 7) (s3') {\Large $\mathbf{F_0}$};
  
  \node[draw,circle,fill=customred] at (0-2, 11) (m1') {\Large $\mathbf{D_0}$};
  \node[draw,circle,fill=customred] at (0-0.5, 10) (m2') {\Large $\mathbf{E_0}$};
  \node[draw,circle,fill=customred] at (0+2, 9) (m3') {\Large $\mathbf{G_0}$};
  \node[draw,circle,fill=customred] at (0+0.5, 8) (m4') {\Large $\mathbf{H_0}$};
  
  \draw[black,line width=3pt] (root') -- (s1');
  \draw[cyan,line width=3pt] (s1') -- (s2');
  \draw[cyan,line width=3pt] (s2') -- (m1');
  \draw[black,line width=3pt] (s2') -- (m2');
  \draw[black,line width=3pt] (s1') -- (s3');
  \draw[black,line width=3pt] (s3') -- (m3');
  \draw[black,line width=3pt] (s3') -- (m4');
  
  \node[font=\huge] at (0+0,-2) (l') {$T_0$};
  
  
  \node[draw,ellipse,fill=gray!70] at (14+0, 0) (root''') {\Large $\mathbf{A_0,A_1}$};
  \node[draw,ellipse,fill=gray!70] at (14+0, 2.5) (s1''') {\Large $\mathbf{B_0,B_1}$};
  \node[draw,circle,fill=gray!70] at (14-1.5, 8.5) (s2''') {\Large $\mathbf{C_0}$};
  \node[draw,circle,fill=gray!70] at (14-1, 4.5) (s22''') {\Large $\mathbf{C_1}$};
  \node[draw,circle,fill=gray!70] at (14+1.5, 6.5) (s3''') {\Large $\mathbf{F_0}$};
  
  \node[draw,ellipse,fill=customred] at (14-2, 10.5) (m1''') {\Large $\mathbf{D_0,D_1}$};
  \node[draw,circle,fill=customred] at (14-0.5, 9) (m2''') {\Large $\mathbf{E_0}$};
  \node[draw,circle,fill=customred] at (14-0.25, 5.5) (m22''') {\Large $\mathbf{E_1}$};
  \node[draw,ellipse,fill=customred] at (14+2, 8.5) (m3''') {\Large $\mathbf{G_0,G_1}$};
  \node[draw,circle,fill=customred] at (14+0.5, 7) (m4''') {\Large $\mathbf{H_0}$};
  
  \draw[black,line width=3pt] (root''') -- (s1''');
  \draw[cyan,line width=3pt] (s1''') -- (s22''');
  \draw[black,line width=3pt] (s1''') -- (s3''');
  \draw[cyan,line width=3pt] (s22''') -- (s2''');
  \draw[black,line width=3pt] (s22''') -- (m22''');
  \draw[cyan,line width=3pt] (s2''') -- (m1''');
  \draw[black,line width=3pt] (s2''') -- (m2''');
  \draw[black,line width=3pt] (s3''') -- (m3''');
  \draw[black,line width=3pt] (s3''') -- (m4''');
  
  \node[font=\huge] at (14+0,-2) (l''') {$T_{0.5}$};
  
  
  \draw[dashed,line width=2.5pt,customblue] (root) to[bend left=0] (root');
  \draw[dashed,line width=2.5pt,customblue] (s1) to[bend left=0] (s1');
  \draw[dashed,line width=2.5pt,customblue] (m1) to[bend right=20] (m1');
  \draw[dashed,line width=2.5pt,customblue] (m3) to[bend left=0] (m3');
  
  \end{tikzpicture}
  }
  \caption{Merge trees $T_0$ and $T_1$ with barycenter $T_{0.5}$. The optimal path mapping between $T_0$ and $T_1$ is illustrated by the dotted lines. Two mapped paths and interpolated path in the geodesic tree are highlighted in cyan. Scalar values and edge lengths can be read from the grid.}
  \label{fig:geodesic_construction}
\end{figure}

Intuitively, for a given $\alpha$, we interpolate the two trees with coefficients $\alpha$ and $(1-\alpha)$.
We first interpolate all mapped paths and move the nodes on these paths such that their relative position on the path remains constant.
Then, we interpolate the inserted/deleted edges from/to length zero.
Note that the problem of contradicting paths described before does not arise here, as only one mapping is considered.
Structurally, the interpolated tree is the supertree induced by the path mapping.
To define scalars, we first interpolate the labels of the matched nodes, i.e.\ the start and end vertices of the matched paths.
Then, we move the nodes on these paths such that their relative position stays the same.
Furthermore, we contract all deleted or inserted edges to $1-\alpha$ or $\alpha$ of their original lengths.
An example is shown in \autoref{fig:geodesic_construction}.

For a formal definition, recall that we say a vertex $v \in V(T_0)$ is present in $M$ (and write $v \in M$) if there is a pair of paths $(p,p') \in M$ such that $v$ is in $p$ (and analogously for vertices of $T_1$).
Furthermore, we assume that $V(T_0)$ and $V(T_1)$ are disjoint and denote the scalar function on the union $V(T_0) \dot{\cup} V(T_1)$ of nodes by $f$. 
Now we define $V(T_\alpha)$ to be the set
\begin{gather*}
\{ (v,v'),(u,u') \mid (v, \dots ,u,v', \dots ,u') \in M \} \\
\cup \{ v_i,u_j \mid (v_0, \dots ,v_k,u_0, \dots ,u_{k'}) \in M, 1 \leq i < k, 1 \leq j < k' \} \\
\cup \{ v \in V(T_0) \mid v \notin M \} \cup \{ v \in V(T_1) \mid v \notin M
\}. 
\end{gather*}
In the example in \autoref{fig:geodesic_construction}, the ellipse nodes form the first set, nodes $C_0,C_1,F_0$ form the second set and $E_0,E_1,H_0$ form the last set.

Next, we define the edge set of $T_\alpha$.
For a pair of mapped paths $(v_1 \dots v_k,u_1 \dots u_{k'}) \in M$ (an example path is highlighted in \autoref{fig:geodesic_construction}), let $s_1,s_2, \dots ,s_{k+k'-4}$ be the sorted union of the inner nodes of the two paths, i.e.\ $f(s_1) \leq f(s_2) \leq \dots \leq f(s_{k+k'-4})$ (in the example, this sequence is $C_0C_1$).
For each such path in $M$, we then include the edges $((v_1,u_1),s_1)$, $(s_{k+k'-4},(v_k,u_{k'}))$ and $(s_i,s_{i+1})$ for each $1 \leq i < k+k'-4$.
Furthermore, we include the edge $(v,v') \in E(T_0)$ if $v \notin M$ and the same way for edges of $T_1$.
In the case where $v'$ is the start or end vertex of a mapped path, we have to replace it by its corresponding node in $V(T_\alpha)$ (the resulting path in $T_\alpha$ is also highlighted in the example).

As a last step, we need to define the scalar function on the new nodes.
For the matched nodes $(v,v')$ (i.e.\ $v,v'$ are the start or end nodes on two mapped paths), we define $f_\alpha((v,v')) = (1-\alpha) \cdot f_0(v) + \alpha \cdot f_1(v')$.
For easier notation, we also write $f_\alpha(v)$ or $f_\alpha(v')$ instead of $f_\alpha((v,v'))$.
For a node $p_i$ on a path $p_1 \dots p_k$ matched to $p'_1 \dots p_{k'}$ with $1 \neq i \neq k$, we define $f_\alpha(p_i) = \frac{f_0(p_i)-f_0(p_1)}{f_0(p_k)-f_0(p_1)} \cdot (f_\alpha(p_k)-f_\alpha(p_1)) + f_\alpha((p_1,p_k))$.
For a deleted node $v \notin M$ with parent $p$, we define $f_\alpha(v) = f_\alpha(p) + (1-\alpha) \cdot (f_0(v) - f_0(p))$.
For an inserted node $v \notin M$ with parent $p$, we define $f_\alpha(v) = f_\alpha(p) + \alpha \cdot (f_0(v) - f_0(p))$.
Note that at least one pair of paths containing the roots of both trees is in the optimal mapping and thus the recursive definition above is well-defined.
Furthermore, the described tree is indeed the result of the first iteration barycenter computation in \autoref{sec:method} when the number of inputs is two.

Based on this definition, we can now show that the merge trees $T_\alpha$ ($0 \leq \alpha \leq 1$) define a geodesic between $T_0$ and $T_1$.
Clearly, the tree $T_\alpha$ can be created from $T_0,T_1$ in linear time.

\textcolor{black}{Recall that, for a metric $d$, a continuous path $P = (T_\alpha)_{0 \leq \alpha \leq 1}$ between two trees $T_0,T_1$ is a geodesic if its length
$$ \mathcal{L}(P) = \sup_{n;0=t_0 \leq t_1 \leq ... \leq t_n=1} \sum_{k=0}^{n-1} d(T_{t_k},T_{t_{k+1}}) $$
is exactly the distance $d(T_1,T_2)$ between $T_1$ and $T_2$.}

Now consider two time points $s,t \in [0,1]$.
With the above definition, we can derive a path mapping $M_{s,t}$ between $T_s$ and $T_t$ from $M$.
We have to make a case distinction on whether the one of the two time points is 0 or 1.

For $0 < s \leq t < 1$, the two trees are structurally the same (only the labels differ).
We define $M_{s,t}$ to be the identity mapping on the edges, which is obviously a valid path mapping.
Thus, $\delta(T_s,T_t) \leq c(M_{s,t})$.
Next, we determine the cost of $M_{s,t}$.

\begin{figure*}
    \centering
    \includegraphics[width=0.16\linewidth]{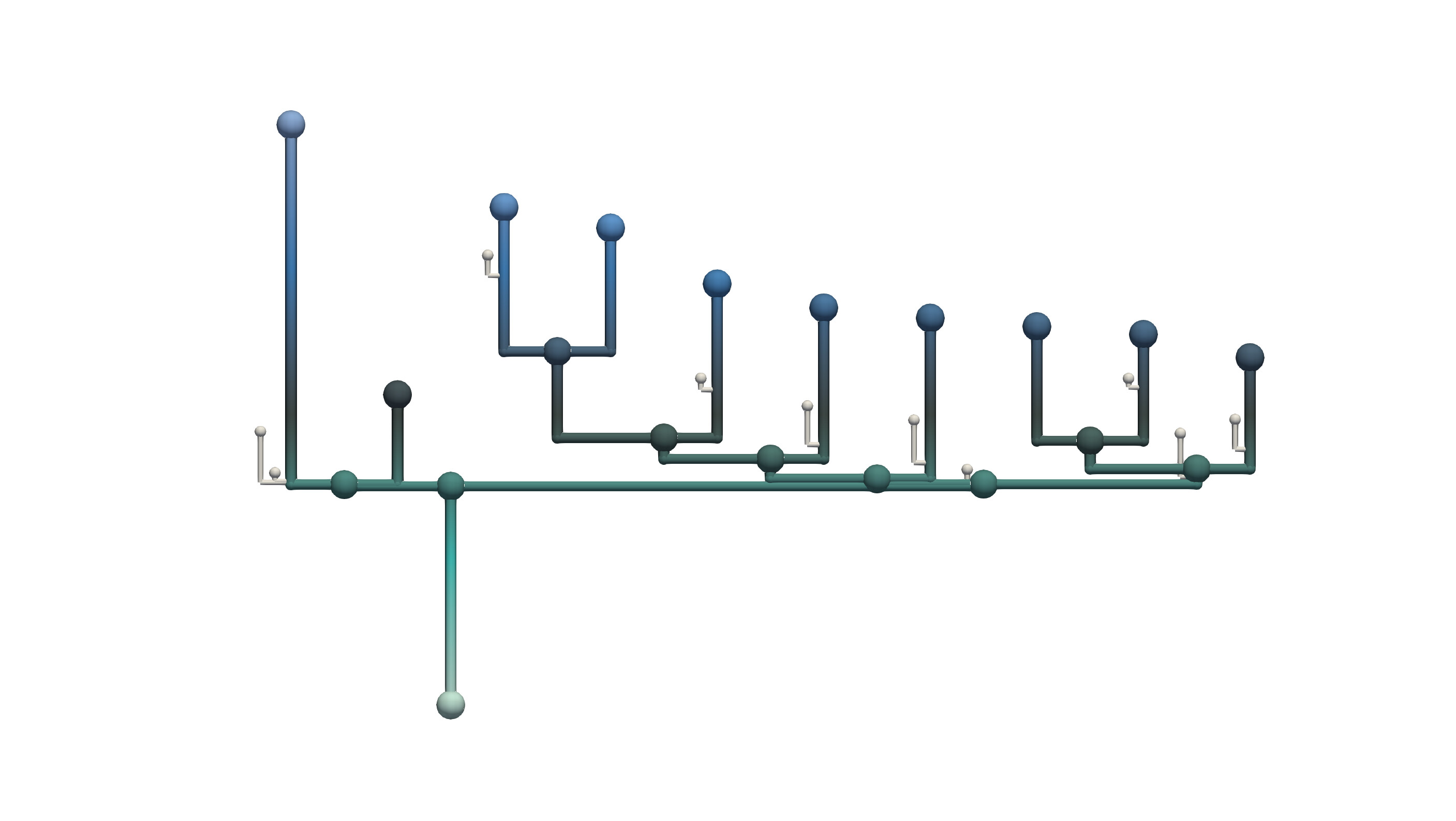}
    \includegraphics[width=0.16\linewidth]{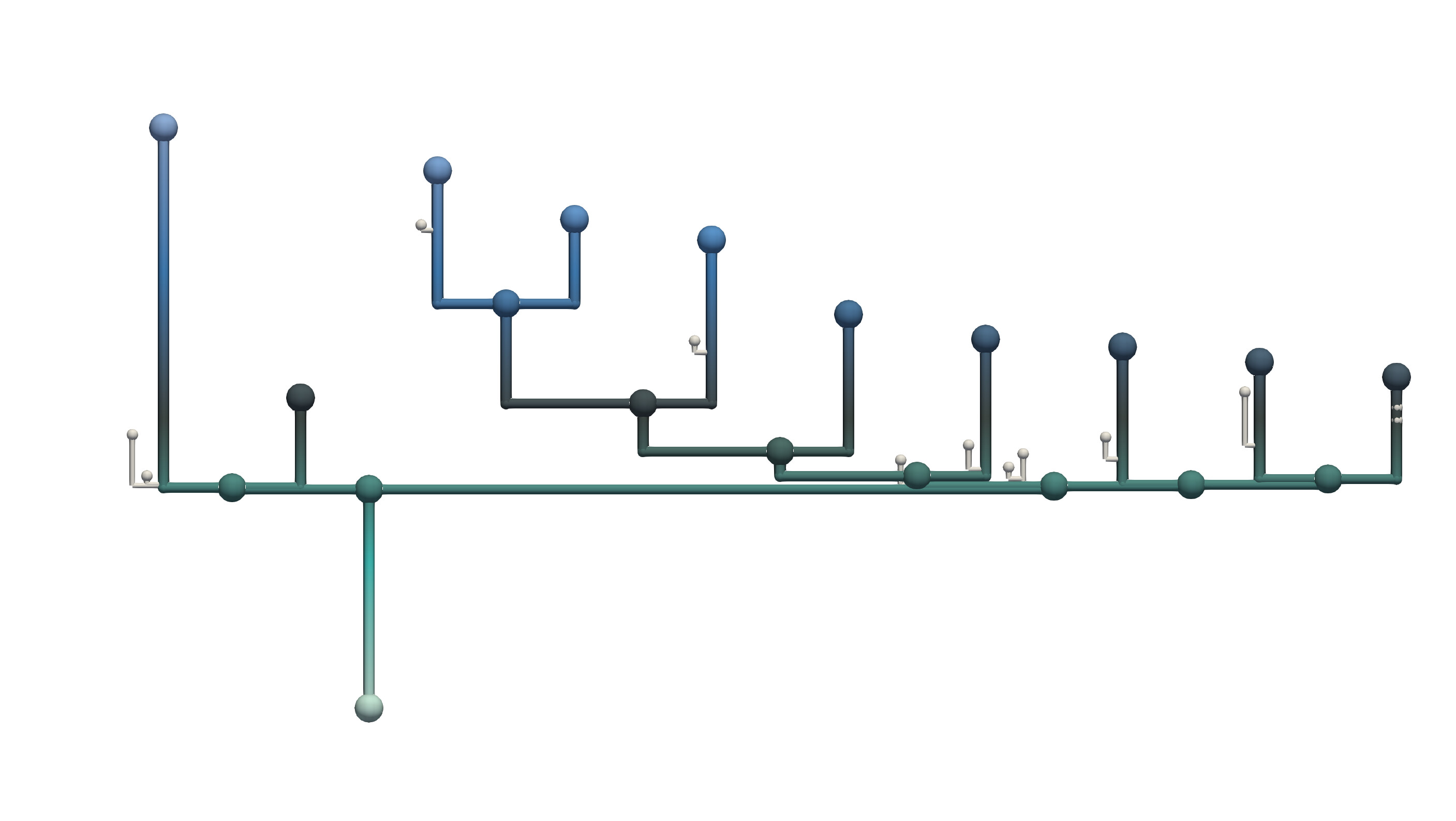}
    \includegraphics[width=0.16\linewidth]{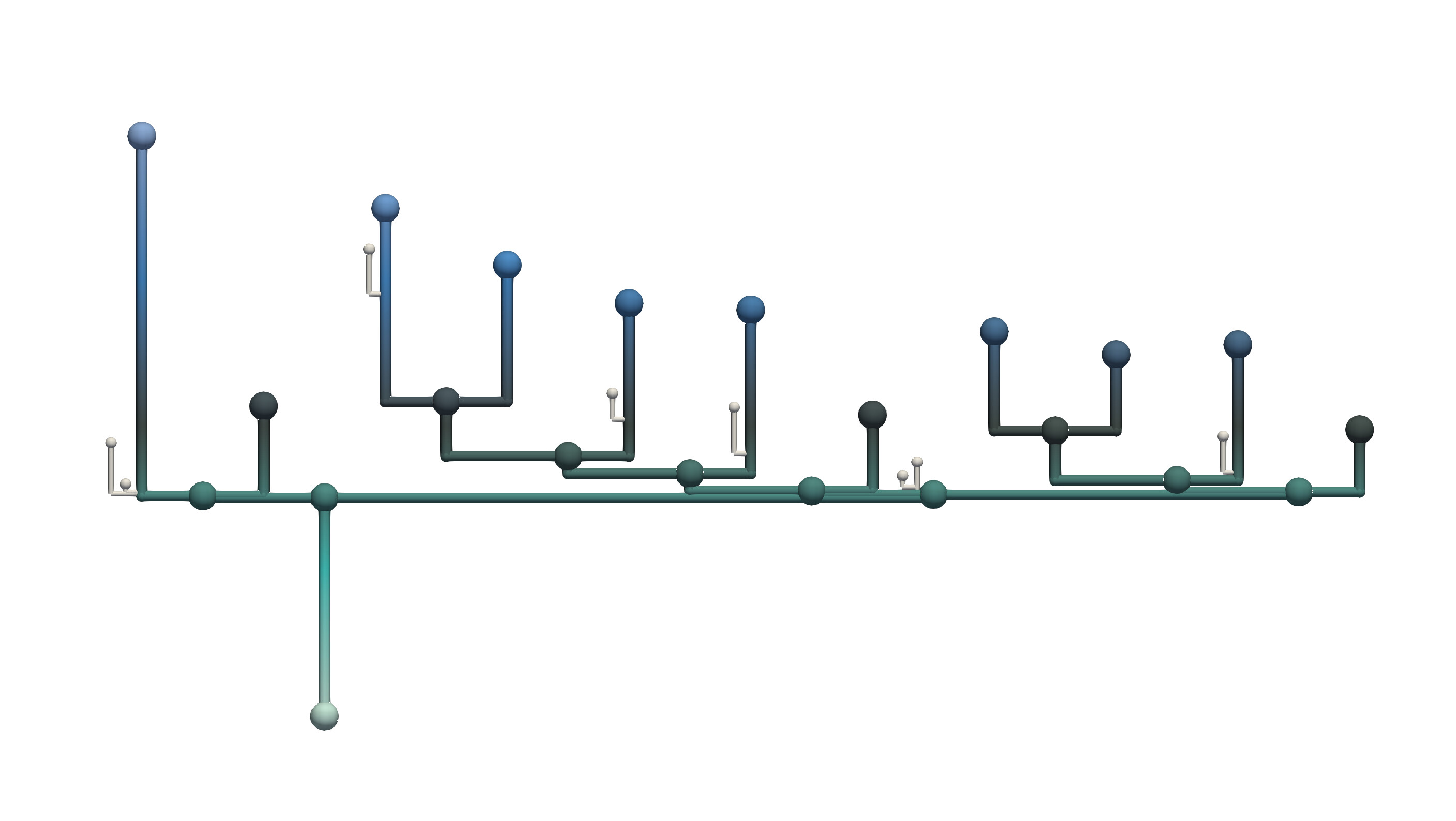}
    \includegraphics[width=0.16\linewidth]{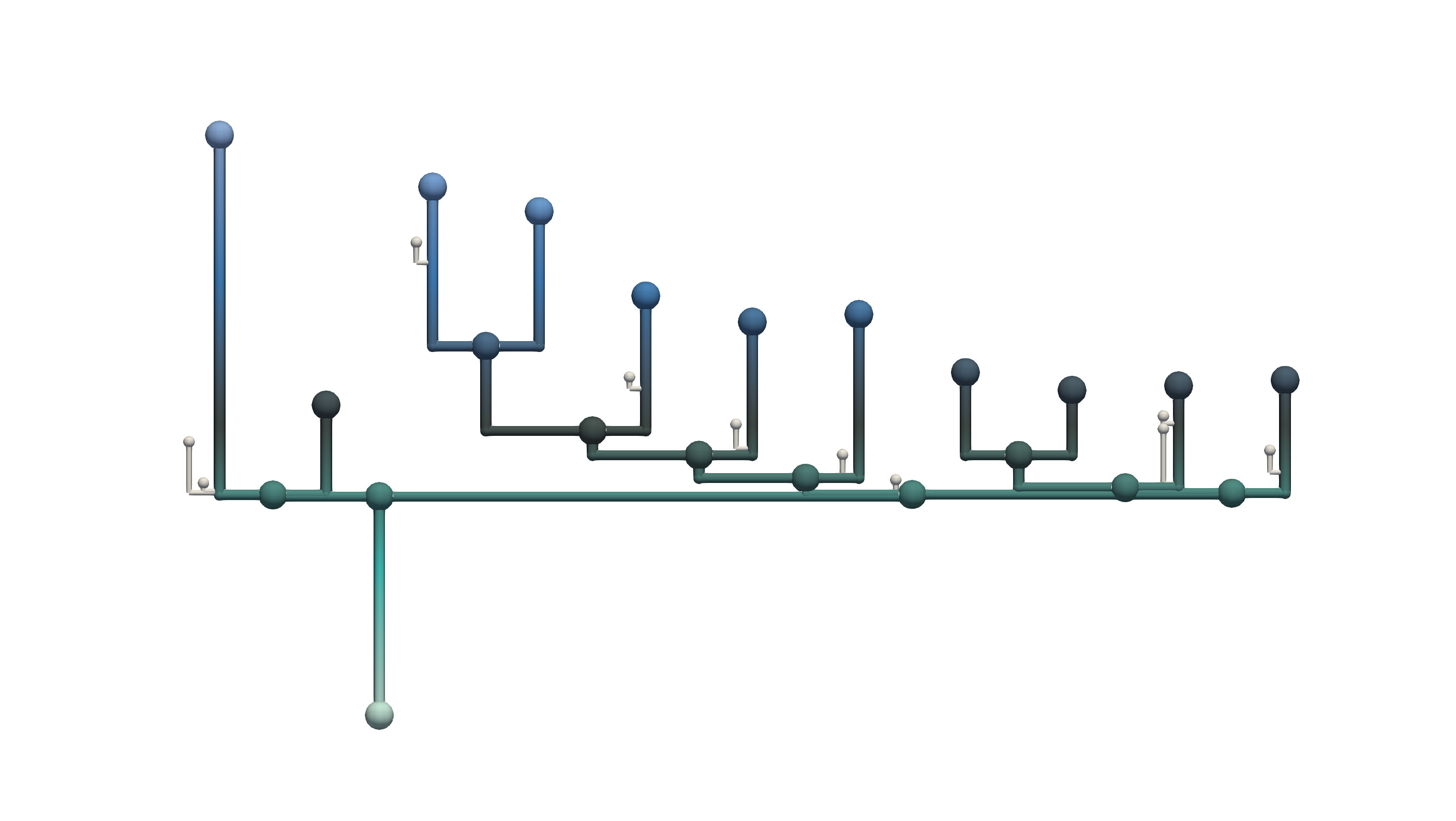}
    \includegraphics[width=0.16\linewidth]{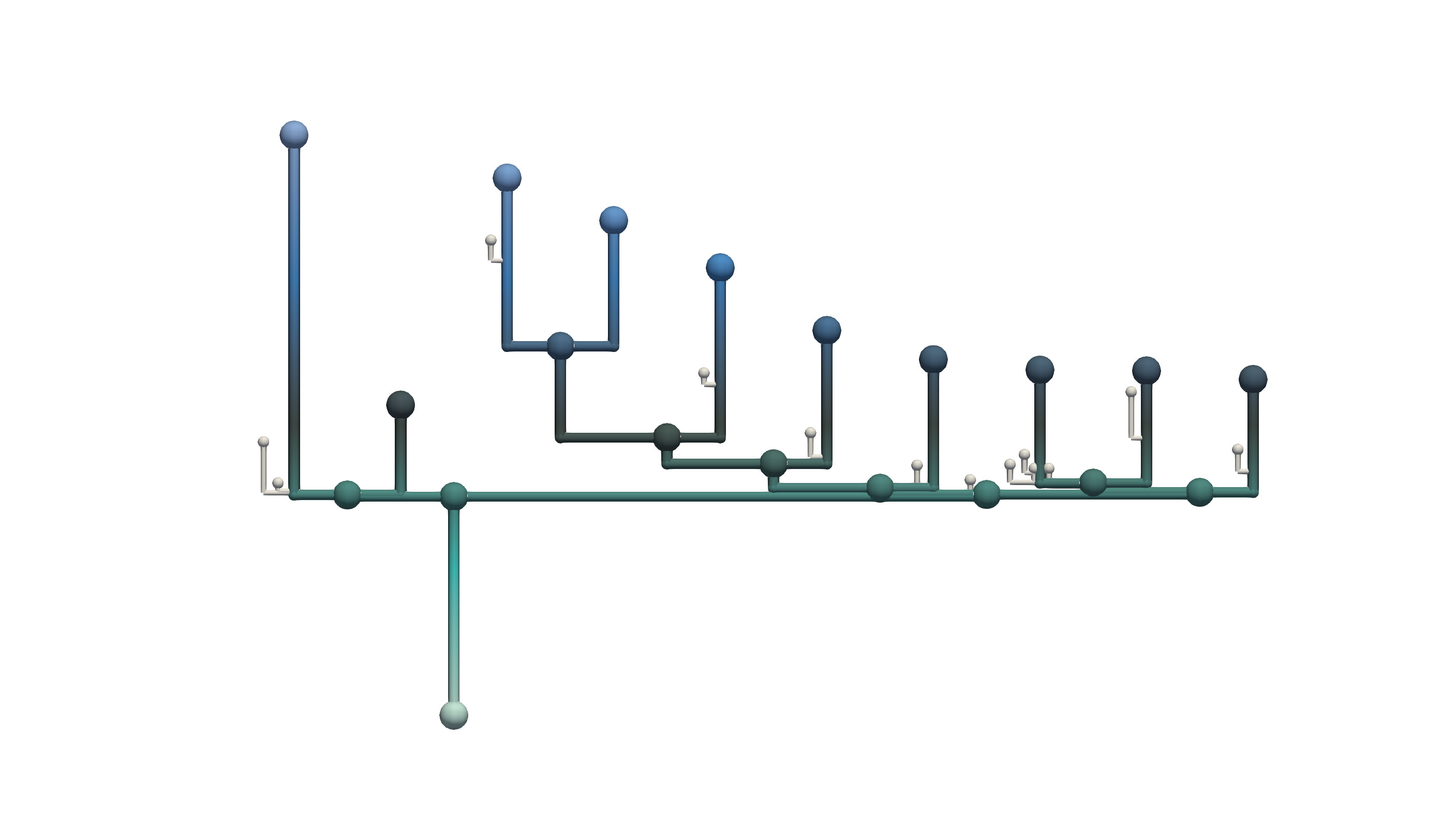}
    \includegraphics[width=0.16\linewidth]{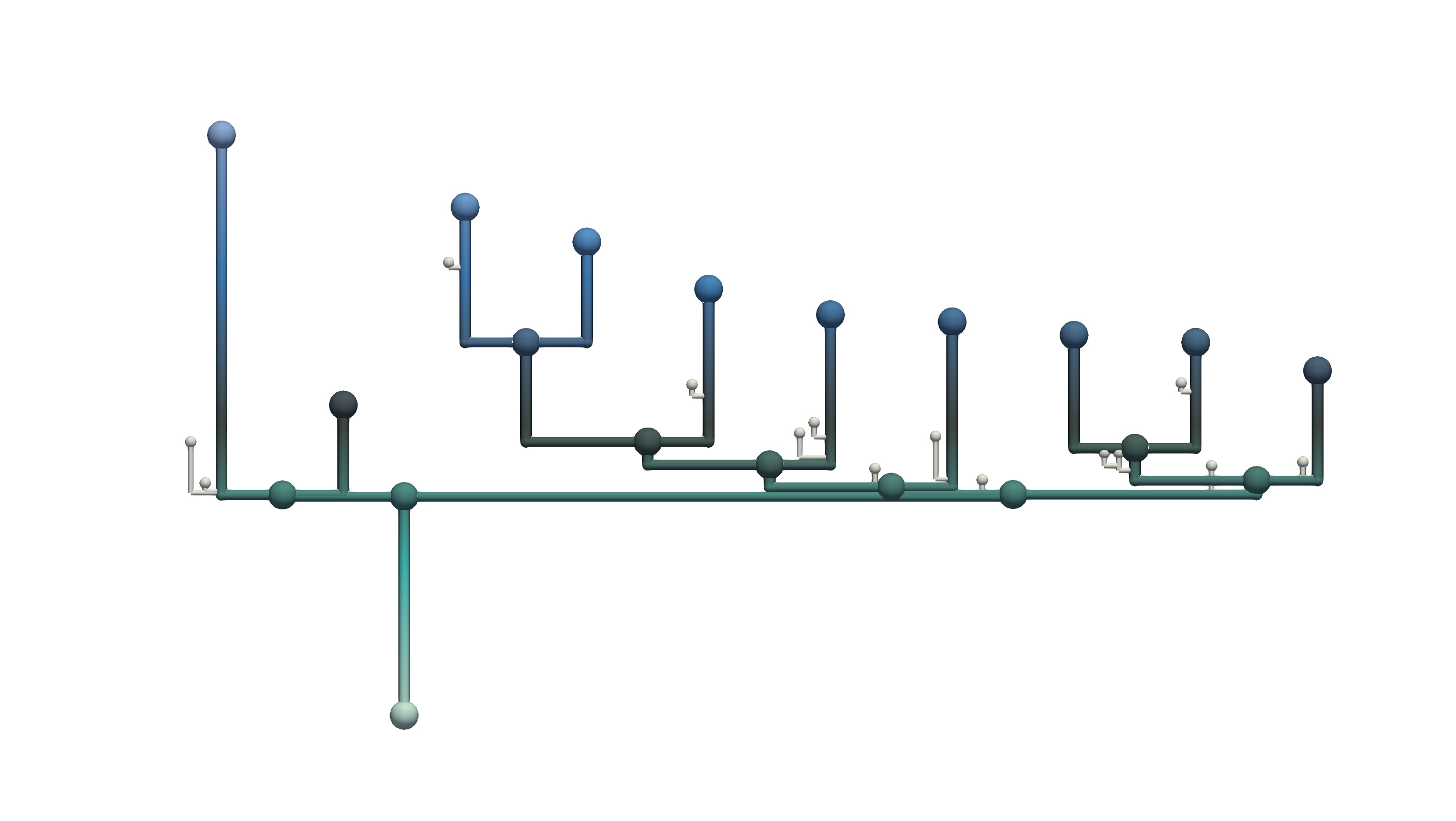}
    
    \includegraphics[width=0.16\linewidth]{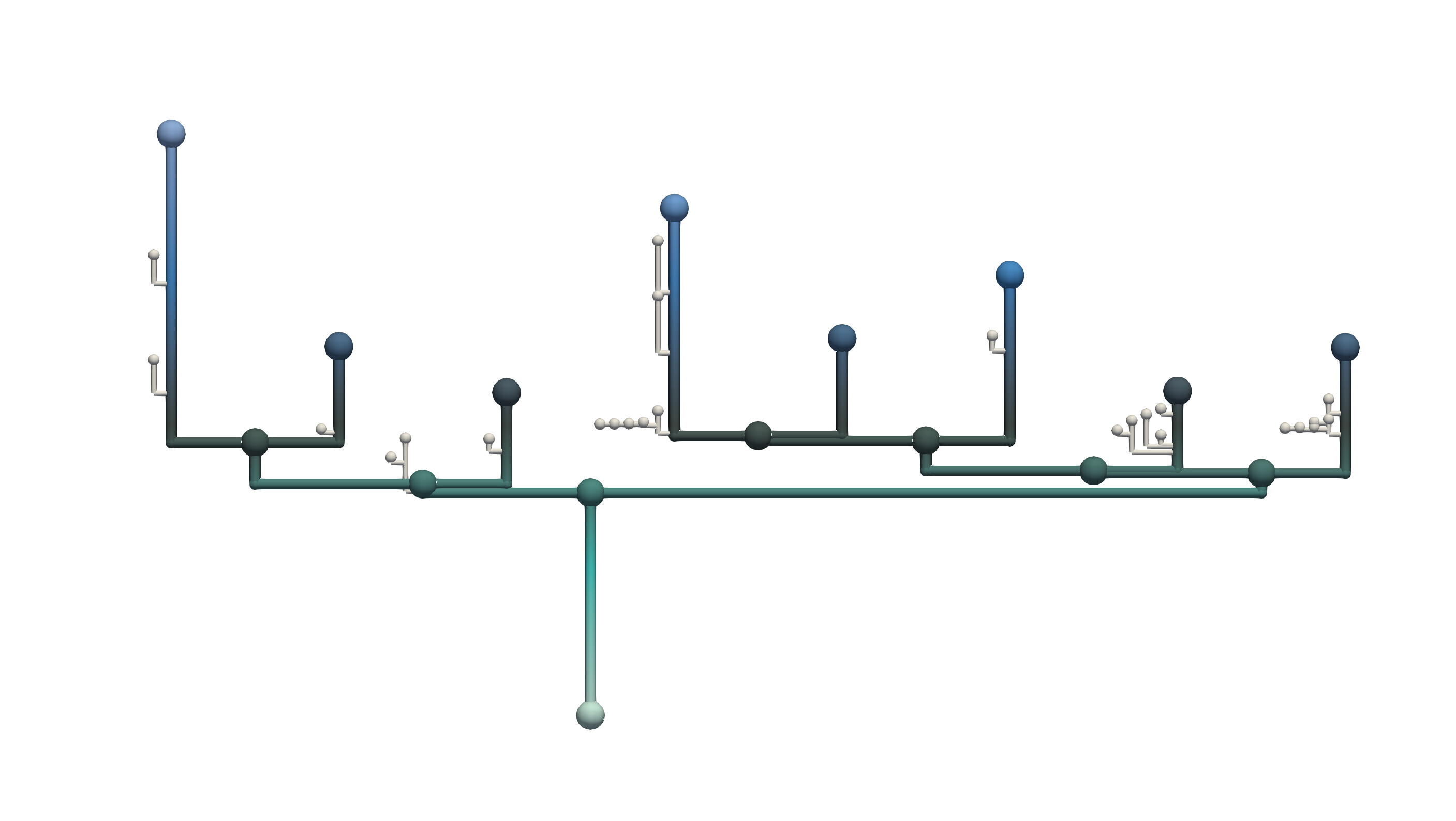}
    \includegraphics[width=0.16\linewidth]{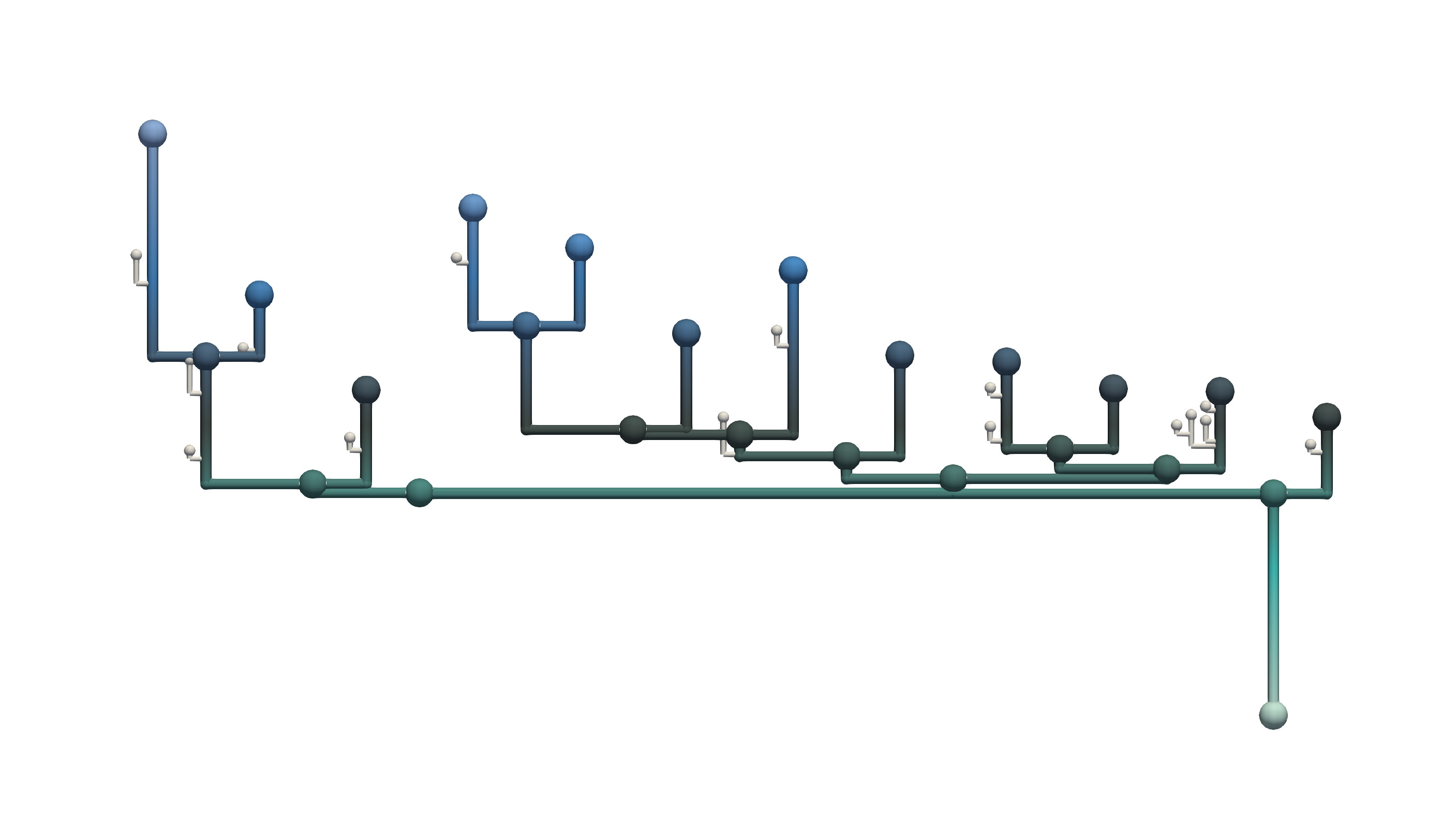}
    \includegraphics[width=0.16\linewidth]{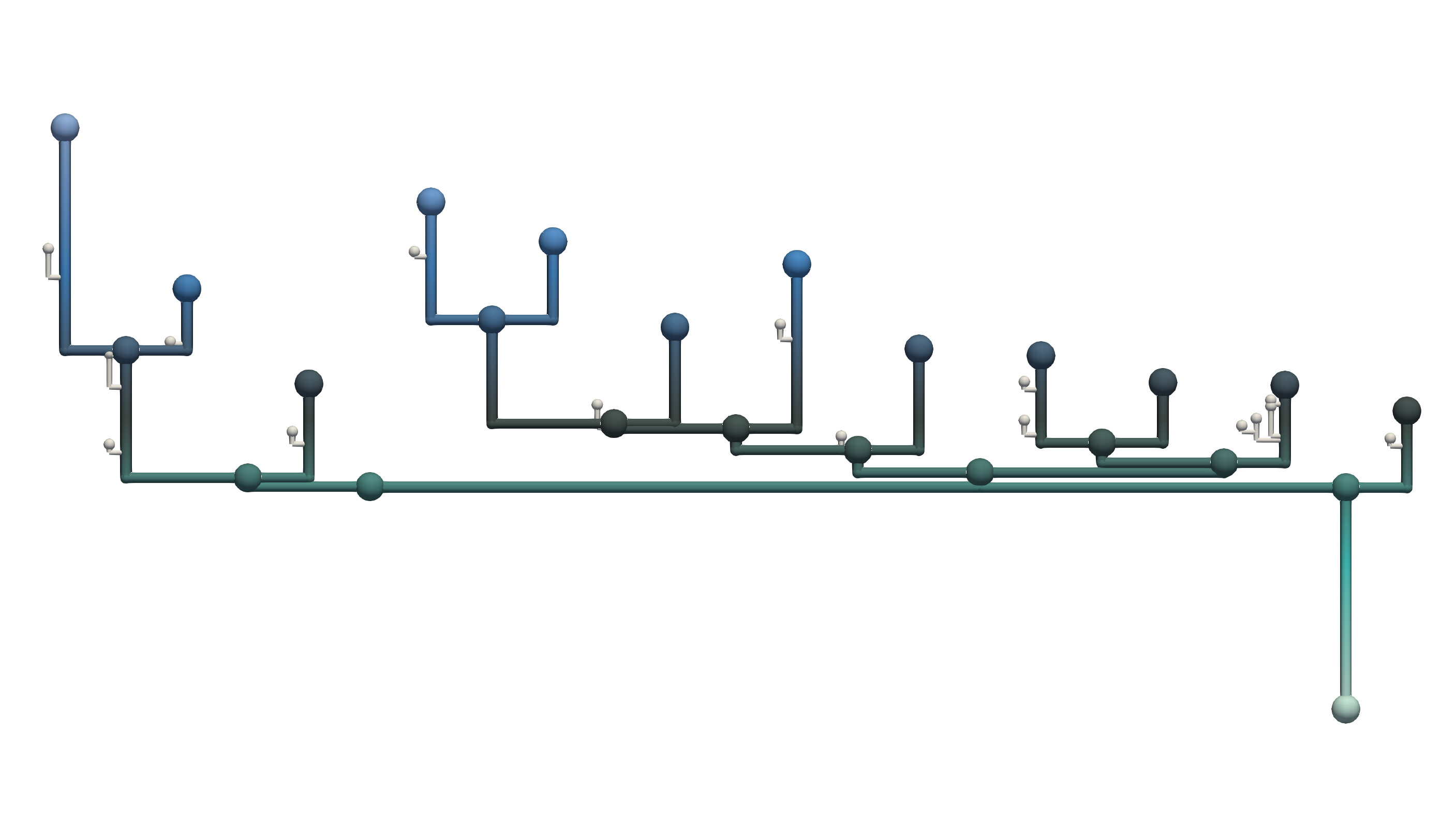}
    \includegraphics[width=0.16\linewidth]{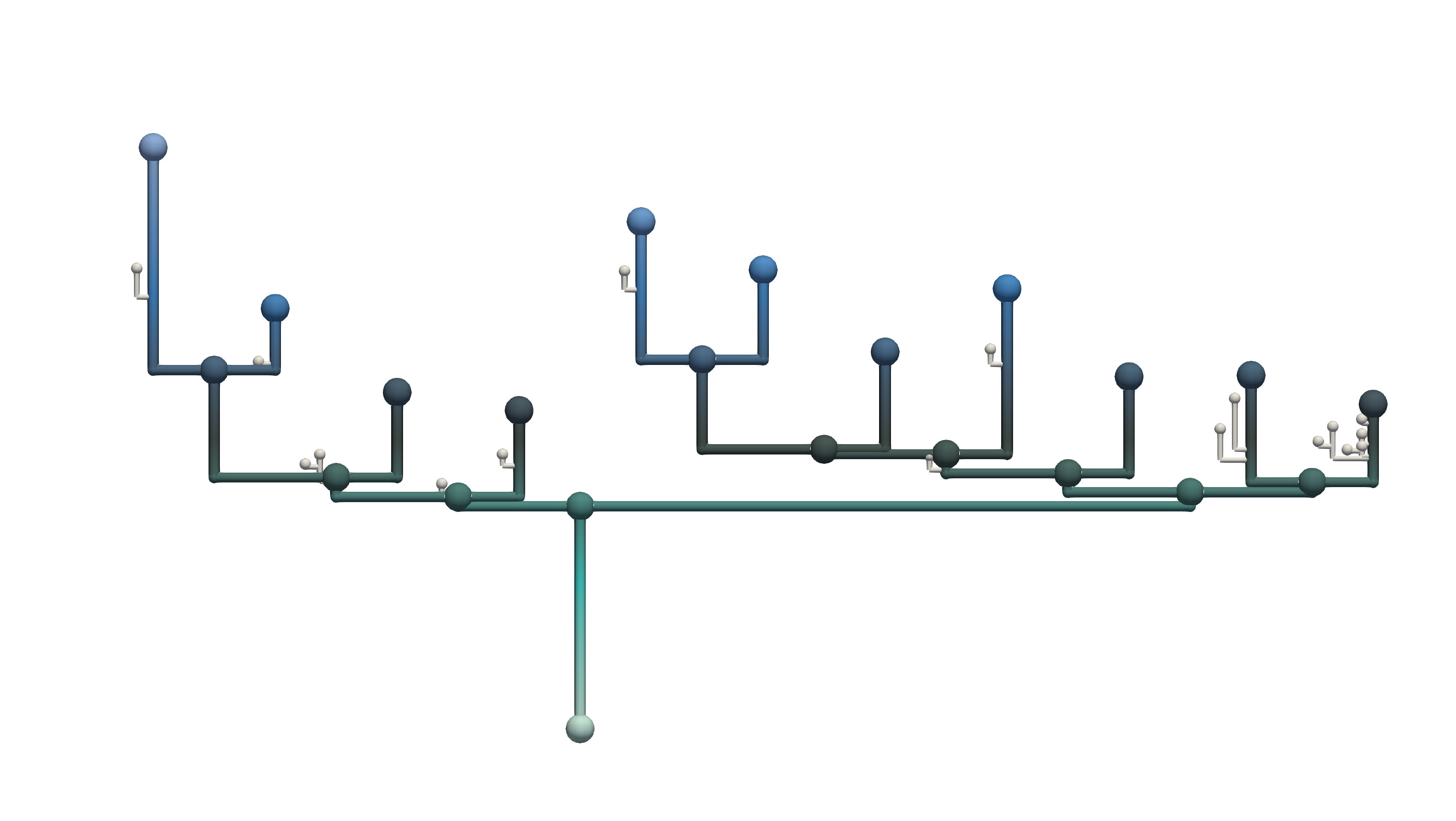}
    \includegraphics[width=0.16\linewidth]{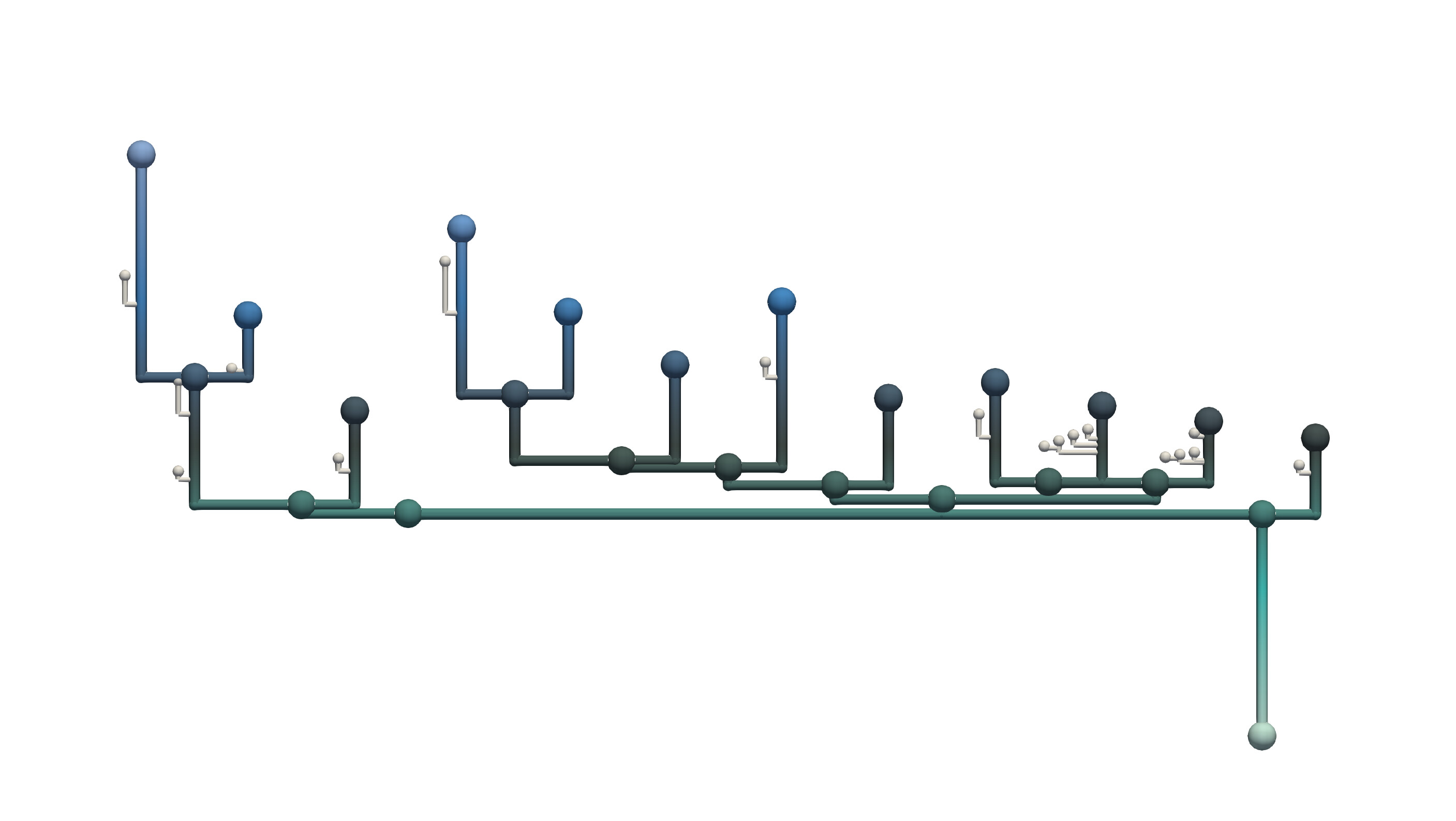}
    \includegraphics[width=0.16\linewidth]{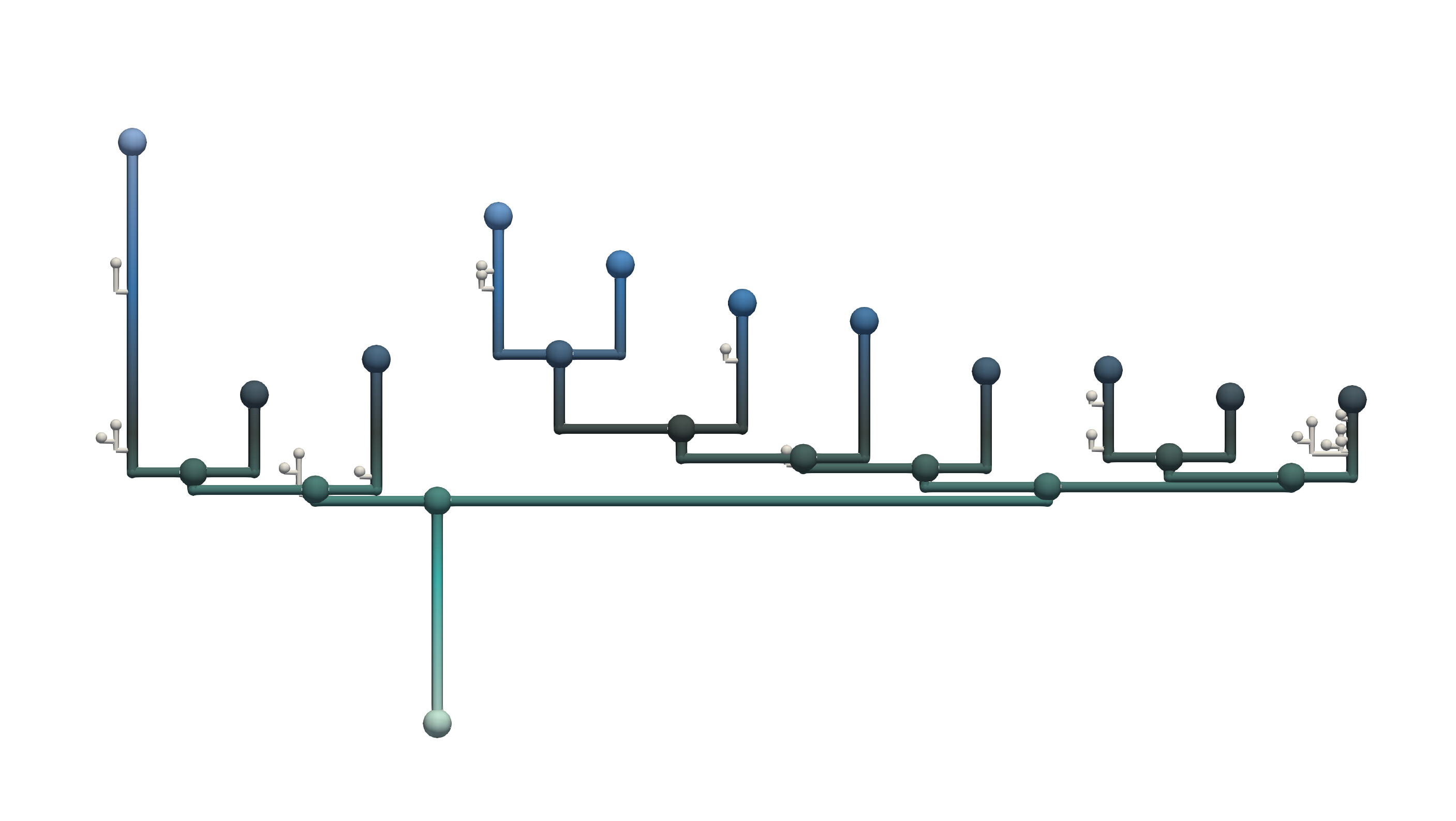}
    
    \caption{All possible barycenter results on the starting vortex ensemble. The top row shows the path mapping barycenters for each of the six initial candidates. The bottom row shows the corresponding Wasserstein barycenters.}
    \label{fig:barycenter_startingvortex_complete}
\end{figure*}

Let $I$ and $D$ be the inserted and deleted edges of $M$.
We have $(e,e) \in M_{s,t}$ for each $e \in I \cup D$.
Each $e \in I$ contributes $c(0,\ell_1(e)) = \ell_1(e)$ in  $c(M)$, whereas they contribute $c(\ell_s(e),\ell_t(e))$ in $c(M_{s,t})$.
By definition, $\ell_s(e) = s \cdot \ell_1(e)$ and $\ell_t(e) = t \cdot \ell_1(e)$ and therefore $c(\ell_s(e),\ell_t(e)) = (t-s) \cdot \ell_1(e)$.
Analogously, each $e \in D$ contributes $c(\ell_0(e),0) = \ell_0(e)$ in  $c(M)$, whereas they contribute $c(\ell_s(e),\ell_t(e))$ in $c(M_{s,t})$.
By definition, $\ell_s(e) = (1-s) \cdot \ell_0(e)$ and $\ell_t(e) = (1-t) \cdot \ell_0(e)$ and therefore $c(\ell_s(e),\ell_t(e)) = (t-s) \cdot \ell_0(e)$.

\begin{figure}
    \centering
    \includegraphics[width=\linewidth]{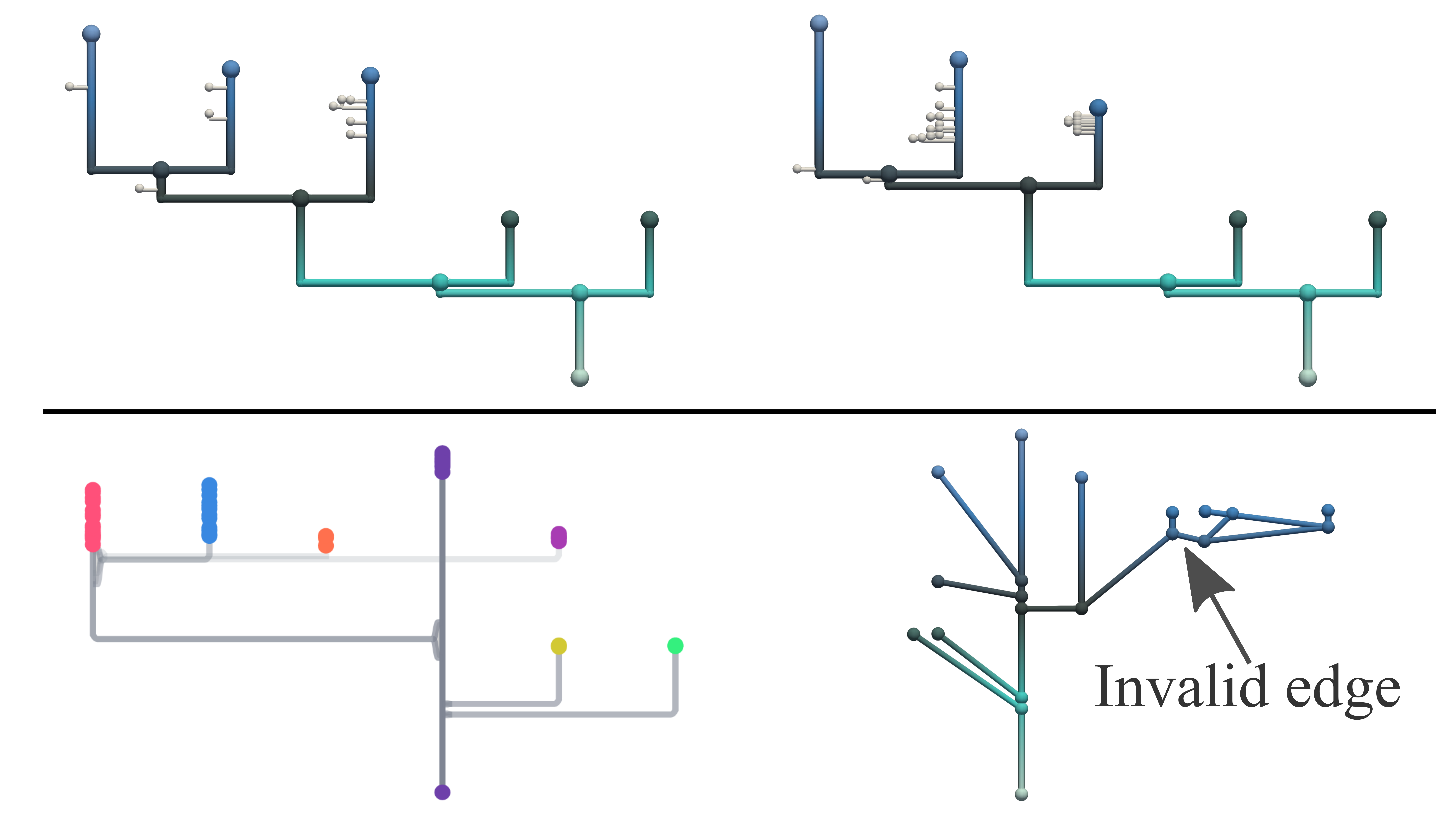}
    \caption{Comparison of merge tree barycenters and contour tree alignments: The top left image shows the path mapping barycenter, the top right image the Wasserstein barycenter. The bottom row shows the fuzzy contour tree on the left and the ParaView rendering of the alignment tree on the right. The latter illustrates the problem of the fuzzy contour tree summarization: the ensemble representative is not a valid merge tree.}
    \label{fig:alignment_heatedCylinder}
\end{figure}

\begin{figure*}
    \centering
    \includegraphics[angle=270,width=170mm]{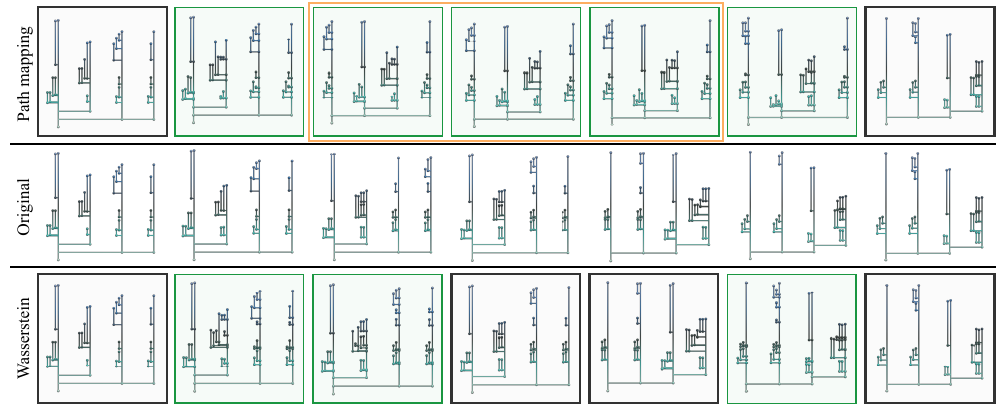}
    \caption{Exemplary time steps of the temporal reduction and reconstruction. The original time series is shown in the middle row. The top row shows the result of the path mapping geodesic, the bottom row of the Wasserstein geodesics. The path mapping reconstruction produces merge trees with a diiferent branch decomposition (to the original series) in the time steps highlighted in yellowm, which is not the case for the Wasserstein geodesics. In particular, the original trees and their Wasserstein reconstructions have the long fork structure as main branch, whereas the path mapping reconstruction has a different one. However, this does not change the fact the path mapping reconstructions are very similar to the original series (when ignoring the branch order).}
    \label{fig:ionization_tempred_2vs4}
\end{figure*}

\begin{table*}[]
  \centering
  \begin{subtable}{\linewidth}
  \centering
  \begin{tabular}{l|c|c|c|c|c|c|c|c|c|c|c|c|c}
  Path mapping distance & TP0 & TP1 & TP2 & TP3 & TP4 & TP5 & TP6 & TP7 & TP8 & TP9 & TP10 & TP11 & TP12 \\
  \hline
  Wasserstein Geodesic & \textbf{0.0} & 0.63 & 0.85 & 1.2 & 1.02 & 0.74 & \textbf{0.0} & 1.59 & 2.59 & 3.54 & 3.09 & 1.6 & \textbf{0.0}\\
  Path mapping Geodesic & \textbf{0.0} & 0.52 & 0.62 & 0.86 & 0.78 & \textbf{0.0} & 0.68 & 0.96 & 0.98 & 1.04 & 0.75 & 0.53 & \textbf{0.0}
  \end{tabular}
  \end{subtable}
  
  \vspace{4pt}
  
  \begin{subtable}{\linewidth}
  \centering
  \begin{tabular}{l|c|c|c|c|c|c|c|c|c|c|c|c|c}
  Wasserstein distance & TP0 & TP1 & TP2 & TP3 & TP4 & TP5 & TP6 & TP7 & TP8 & TP9 & TP10 & TP11 & TP12 \\
  \hline
  Wasserstein Geodesic & \textbf{0.0} & 0.06 & 0.17 & 0.18 & 0.11 & 0.04 & \textbf{0.0} & 1.34 & 0.87 & 0.49 & 0.22 & 0.06 & \textbf{0.0}\\
  Path mapping Geodesic & \textbf{0.0} & 0.26 & 0.37 & 0.09 & 0.06 & \textbf{0.0} & 0.01 & 0.03 & 0.04 & 0.05 & 0.03 & 0.02 & \textbf{0.0}
  \end{tabular}
  \end{subtable}
  \caption{Comparison of the temporal reduction results based on path mapping geodesics and Wasserstein geodesics. The first table shows the path mapping distance between the original and reconstructed merge trees for each time \textcolor{black}{point} and both methods. The second table depicts the Wasserstein distances. Keyframes are again highlighted in bold.}
  \label{tab:distance}
\end{table*}

Now consider the rest of the edges in $T_s$ and $T_t$.
They are constructed from a mapped path as described above.
A pair of mapped paths (we assume leaf-to-root direction in a split tree) $(p_1...p_k,p'_1...p'_{k'}) \in M$ contributes
$$||f_0(p_1) - f_0(p_k)| - |f_1(p'_1) - f_1(p'_{k'})||$$
$$= f_0(p_1) - f_0(p_k) - (f_1(p'_1) - f_1(p'_{k'}))$$
to $c(M)$.
Instead of $p_1...p_k$ and $p'_1...p'_{k'}$ in $T_0,T_1$, we have in $T_\alpha$ the nodes
$$s_0 := (p_1,p'_1), s_1,s_2,...,s_{k+k'-4},s_{k+k'-3} := (p_k,p'_{k'})$$
as well as the edges $(s_i,s_{i+1})$ for each $0 \leq i \leq k+k'-4$.

Each mapped edge in $M_{s,t}$ contributes 
$$||f_s(s_i) - f_s(s_{i+1})| -|f_t(s_i) - f_t(s_{i+1})||$$
to $c(M_{s,t})$.
Thus, the whole path contributes
$$\sum_{0 \leq i \leq k+k'-4} ||f_s(s_i) - f_s(s_{i+1})| -|f_s(s_i) - f_s(s_{i+1})||.$$
Note that the whole path and thus each single edge gets either shorter or longer.
So we either have:
\begin{itemize}
    \item $|f_s(s_0) - f_s(s_{k+k'-3})| > |f_t(s_0) - f_t(s_{k+k'-3})|$ and $|f_s(s_i) - f_s(s_{i+1})| > |f_t(s_i) - f_t(s_{i+1})|$ for each $i$ or
    \item $|f_s(s_0) - f_s(s_{k+k'-3})| < |f_t(s_0) - f_t(s_{k+k'-3})|$ and $|f_s(s_i) - f_s(s_{i+1})| < |f_t(s_i) - f_t(s_{i+1})|$ for each $i$.
\end{itemize}
W.l.o.g.\ we have $|f_s(s_i) - f_s(s_{i+1})| > |f_t(s_i) - f_t(s_{i+1})|$ for each $i$.
In total, we get for the cost of mapping the whole path:
$$\sum_{0 \leq i \leq k+k'-4} ||f_s(s_i) - f_s(s_{i+1})| -|f_s(s_i) - f_s(s_{i+1})||$$
$$ = \sum_{0 \leq i \leq k+k'-4} (f_s(s_i) - f_s(s_{i+1})) - (f_t(s_i) - f_t(s_{i+1}))$$
$$ = \sum_{0 \leq i \leq k+k'-4} (f_s(s_i) - f_s(s_{i+1})) - \sum_{0 \leq i \leq k+k'-4} (f_t(s_i) - f_t(s_{i+1}))$$
$$ = f_s(s_0) - f_s(s_{k+k'-3}) - (f_t(s_0) - f_t(s_{k+k'-3})) $$
$$ = f_s(p_1) - f_s(p_k) - (f_t(p'_1) - f_t(p'_{k'})) $$
$$ = (1-s)f_0(p_1)+s f_1(p'_1) - ((1-s)f_0(p_k)+s f_1(p'_{k'}))$$
$$ - ((1-t)f_0(p_1)+t f_1(p'_1) - ((1-t)f_0(p_k)+t f_1(p'_{k'}))) $$
$$ = (1-s)(f_0(p_1) - f_0(p_k)) + s (f_1(p'_1) - f_1(p'_{k'}))$$
$$ - ((1-t)(f_0(p_1) - f_0(p_k)) + t (f_1(p'_1) - f_1(p'_{k'}))) $$
$$ = (t-s)(f_0(p_1) - f_0(p_k)) + (s-t) (f_1(p'_1) - f_1(p'_{k'})) $$
$$ = (t-s)(f_0(p_1) - f_0(p_k)) - (t-s) (f_1(p'_1) - f_1(p'_{k'})). $$
In total, we get that for each deleted or inserted edge as well as each mapped path that contributes $x$ to $c(M)$ contributes $(t-s)x$ to $c(M_{s,t})$.
Thus, $c(M_{s,t}) = (t-s)c(M)$.

Now consider the case where $0 = s < t$.
Since $T_t$ is structurally a supertree of $T_0$, we can define the mapping $M_{0,t}$ to be the embedding from $T_0$ in $T_t$.
Consider some inserted (in $M$) edge $e \in  I$.
Since $e \notin E(T_0)$, it is also inserted in the embedding $M_{0,t}$.
Thus, it contributes $c(0,\ell_t(e)) = t \cdot \ell_1(e) = (t-s) \cdot \ell_1(e)$ to $M_{0,t}$, whereas it contributes $\ell_1(e)$ to $M$.
A deleted (in $M$) edge $e \in D$ is mapped to itself in the embedding $M_{0,t}$.
Thus, it contributes $c(\ell_0(e),\ell_t(e)) = \ell_0(e) - (1-t) \cdot \ell_0(e) = t \cdot \ell_0(e) = (t-s) \cdot \ell_0(e)$ to $M_{0,t}$, whereas it contributes $\ell_0(e)$ to $M$.

For mapped paths, all arguments are analogous to the previous case.
Thus, in total we again have that for each deleted or inserted edge as well as each mapped path that contributes $x$ to $c(M)$ contributes $(t-s)x$ to $c(M_{s,t})$, and therefore $c(M_{s,t}) = (t-s)c(M)$.
The same holds for the case where $s < t = 1$ with analogous arguments.

\medskip
So we have $c(M_{s,t}) = (t-s)c(M)$ for any two time points $0 \leq s < t \leq 1$.
From this, we can conclude that $\delta(T_s,T_t) \leq (t-s) \delta(T_0,T_1)$.
We can now show that for $P = (T_\alpha)_{\alpha \in [0,1]}$, it holds that $\mathcal{L}(P) = \delta(T_0,T_1)$.

Using the metric property of $\delta$, we know that for any $0 \leq s < t \leq 1$:
$$ \delta(T_0,T_1) \leq \delta(T_0,T_s) + \delta(T_s,T_t) + \delta(T_t,T_1) $$
$$ \leq ((s-0) + (t-s) + (1-t)) \delta(T_0,T_1) = \delta(T_0,T_1). $$
We can conclude that $\delta(T_s,T_t) = (t-s) \delta(T_0,T_1)$ and for any subdivision $0=t_0 < t_1 < \dots < t_n=1$ of $P$, we have:
$$ \sum_{k=0}^{n-1} \delta(T_{t_k},T_{t_{k+1}}) = \delta(T_0,T_1). $$
Thus, for the length of $P$ we get
$$\mathcal{L}(P) = \sup_{n,0=t_0 < t_1 < \dots < t_n=1} \sum_{k=0}^{n-1} \delta(T_{t_k},T_{t_{k+1}}) = \delta(T_0,T_1).$$

\section{Starting Vortex Barycenters}
\label{sec:startingvortex_app}

We now provide further screenshots of the barycenters computed on the starting vortex ensemble.
\autoref{fig:barycenter_startingvortex_complete} shows the barycenter merge trees for each possible initial candidate and both methods.
For five out of six initial candidates, the Wasserstein barycenter contains two fork structures of high persistence, which is not the case in the member trees (see \autoref{fig:barycenter_startingvortex}), whereas only one contains a long, non-forking edge.
In contrast, all six path mapping barycenters are a good summary of the ensemble.

\section{Comparison to contour tree alignments}
\label{sec:heatedCylinder_app}

Next, we quickly illustrate the advantages of path mapping and Wasserstein barycenters over contour tree alignments.
We computed barycenter merge trees, the contour tree alignment and the fuzzy contour tree layout for an ensemble consisting of a fixed time steps (in the late phase of the simulation, see \autoref{fig:barycenter_heatedCylinder}) from different runs of the heated cylinder dataset.
We applied topological simplification with a threshold of 2\% of the scalar range.
\autoref{fig:alignment_heatedCylinder} shows both barycenters, the fuzzy contour tree rendering (see~\cite{LohfinkWLWG20} for details) as well as a ParaView rendering of the alignment tree.
While the branch decomposition layout of the fuzzy contour tree summarizes the ensemble well, the ParaView rendering reveals that the alignment tree is not a valid merge tree.
This is due to a different averaging technique based on the nodes instead of arcs, branches or paths.
It is therefore harder to use for further analysis tasks.

\section{Temporal Reduction}
\label{sec:tempred_app}

In this section, we provide more detailed results for the temporal reduction on the ionization front time series.
In \autoref{sec:experiments}, we compared the reconstructed series of the path mapping geodesics and the Wasserstein geodesics with three key frames for both methods.
The quantitative comparison in terms of actual distances between the original and reconstructed trees are given in~\autoref{tab:distance}.

Furthermore, the path mapping geodesics yield good reconstructions already for two key frames, since it can compute meaningful mappings even between the first and last time step.
The Wasserstein geodesics need four key frames to produce a good reconstruction, since it can not map the first tree to the last one in a meaningful manner.
It therefore needs the time step right before the maximum swap to correctly interpolate the first half of the series and the time step right after the maximum swap to correctly interpolate the second half.
We illustrate this behavior on example time steps in \autoref{fig:ionization_tempred_2vs4}.
The path mapping reconstruction with two key frames is of similar quality to the one with three key frames (rated on a purely visual basis), whereas the Wasserstein reconstruction is significantly improved with four keyframes (visually bad interpolation as highlighted in red in in \autoref{fig:tempred_ionization} do no longer happen).

\bibliographystyle{abbrv-doi-hyperref}

\bibliography{paper}

\makeatletter\@input{xx.tex}\makeatother

%% file: wassersteinDistance.tex
\subsection{Wasserstein Distance and Barycenters}

\input{notations}

Next, we formalize the Wasserstein distance between merge trees as well as Wasserstein barycenters and the algorithm computing them. 
Pont et al. \cite{pont_vis21} introduced a metric between BDTs, formulated as a generalization of the celebrated Wasserstein distance between persistence diagrams \cite{edelsbrunner09}, which we briefly recap here.
Let us first consider the simple case where in both BDTs to compare (denoted $\branchtree_1 := \branchtree(\mathbb{X}_1,f_1)$ and $\branchtree_2 := \branchtree(\mathbb{X}_2,f_2)$), all nodes are direct children of the root.  
This situation corresponds to the classical formulation of the Wasserstein distance between persistence diagrams.
\textcolor{MarkColor}{Each branch in $\branchtree_1$ and $\branchtree_2$ can be represented 
as a 2D point by using its birth value as X coordinate and its death value as Y coordinate.
This yields, for each persistence diagram, a 2D point cloud in the so-called birth/death plane. In this representation, all the branches appear above the diagonal (since a feature dies only after it was created).}

To measure the distance between such BDTs $\branchtree_1$ and $\branchtree_2$, a first pre-processing  step  consists in augmenting each BDT with \textcolor{MarkColor}{projections $\projection(b)$ for all off-diagonal points $b$}
in the other BDT:
\[
  \nonumber
 \branchtree'_1 = \branchtree_1 \cup \{
\projection(b_2) ~ | ~ b_2 \in \branchtree_2
\}, \;
\nonumber
  \branchtree'_2 = \branchtree_2 \cup \{
\projection(b_1) ~ | ~ b_1 \in \branchtree_1
\},
\]

\noindent
where $\projection(b) = (\frac{x+y}{2},\frac{x+y}{2})$ stands for the diagonal projection \textcolor{MarkColor}{(i.e.\ the closest point on the the diagonal)} of the off-diagonal point $b = (x, y)$.
Intuitively, this augmentation phase inserts dummy features in the BDT (with zero persistence, along the diagonal), hence preserving the topological information.
This augmentation guarantees that the two BDTs now have the same number of points ($| \branchtree'_1| = |\branchtree'_2|$), which facilitates the evaluation of their distance, as described next.

Given two points $b_1 = (x_1, y_1) \in \branchtree'_1$ and $b_2 = (x_2, y_2) \in \branchtree'_2$, the ground distance $\pointMetric_2$ in the 2D birth/death space is given by:
\begin{eqnarray}
  \nonumber
\label{eq_pointMetric}
\pointMetric_{2}(b_1, b_2) = (|x_2 - x_1|^2 + |y_2 - y_1|^2)^{1/2}
= \| b_1 - b_2
\|_2.
\end{eqnarray}

\noindent
By convention, $\pointMetric_2(b_1, b_2)$ is set to zero between diagonal points ($x_1 = y_1$ and $x_2 = y_2$). Then, the $L^2$-Wasserstein distance $\wassersteinTree$ is:
\begin{eqnarray}
\label{eq_wasserstein}
\wassersteinTree(\branchtree_1, \branchtree_2)
=
\label{eq_wasserstein_mapping}
&
\hspace{-.15cm}
\underset{\phi \in \Phi}\min
 \hspace{-.15cm}
&
\big(
\sum_{b_1 \in \branchtree'_1}
\pointMetric_2(b_1,
\phi(b_1))^2\big)^{1/2},
\end{eqnarray}

\noindent
where $\Phi$ is the set of all possible assignments $\phi$ mapping a point $b_1 \in \branchtree'_1$ to a point $b_2 \in \branchtree'_2$ (possibly its diagonal projection, indicating the destruction of the corresponding feature).

To account for the general case where the BDTs have an arbitrary structure, one simply needs to update the above formulation by considering a smaller search space of possible assignments, 
noted $\Phi' \subseteq \Phi$, constrained to describe (rooted) partial isomorphisms \cite{pont_vis21} between $\branchtree_1$ and $\branchtree_2$.
Intuitively, $\wassersteinTree$ can be understood as a variant of  the Wasserstein distance between persistence diagrams,
which takes into account the structures of the BDTs when evaluating candidate assignments in the optimization of \autoref{eq_wasserstein}.
In the remainder, we note $\branchtreeSpace$ the metric space induced by the Wasserstein metric between BDTs.

Once distances for BDTs are available, the notion of \emph{barycenter} can be introduced.
Given a set $\branchtreeSet = \{\branchtree_1, \dots, \branchtree_\ensembleSize\}$,
\textcolor{MarkColor}{let $\frechetEnergy(\branchtree)=\sum_{i = 1}^{\ensembleSize}\wassersteinTree(\branchtree,\branchtree_i\big)^2$ be the Fr\'echet energy of a candidate BDT $\branchtree \in \branchtreeSpace$.}

\noindent
Then a BDT $\branchtree^*  \in \branchtreeSpace$
which minimizes $\frechetEnergy$ is called a \emph{Wasserstein barycenter} of the set $\branchtreeSet$ (or its Fr\'echet mean under the metric $\wassersteinTree$).

$\frechetEnergy$
can be optimized with an iterative algorithm \cite{pont_vis21}, alternating \emph{assignment} and \emph{update} phases. The algorithm is reminiscent of Lloyd's relaxation algorithm~\cite{lloyd82}.
Specifically, given a candidate barycenter $\branchtree$, the assignment step consists in computing the optimal assignments $\phi_i$ (w.r.t. \autoref{eq_wasserstein}) between $\branchtree$ and each input BDT $\branchtree_i$.
Next, the update step aims at optimizing
$\frechetEnergy$
under the current set of assignments $\phi_i$, which is achieved by moving, in the birth/death plane, each branch $b$ of $\branchtree$ to the arithmetic mean of its matched branches $\phi_i(b)$.
This assignment/update sequence is then iterated, decreasing the Fr\'echet energy at each iteration.
\textcolor{MarkColor}{Note that this procedure is not guaranteed to converge to a global minimum, i.e.\ the Fr\'echet energy is not necessarily globally minimized.}
Iteration is stopped when the Fr\'echet energy change is less than 1\% between two steps.
\textcolor{MarkColor}{The initial candidate is chosen as a random member of $\branchtreeSet$, see~\cite{pont_vis21} for details.
This choice might influence which local minimum is found.}

To ensure that the interpolated barycenter trees are indeed structurally valid BDTs (i.e.\ invertible into a valid merge trees), Pont et al. introduce a local normalization \cite{pont_vis21} in a pre-processing step.
Specifically, a branch $b = (x,y)$ with parent branch $b' = (x',y')$ is valid/invertible if $[x, y] \subseteq [x', y']$.
Thus, the persistence of each branch $b_i \in \branchtree_i$ is normalized with regard to that of its parent $b'_i \in \branchtree_i$, displacing $b_i$ in the 2D birth/death space such that its position is given relatively to the scalar range of $b'_i$.
After this pre-normalization of the input trees, any interpolated barycenter BDT is reverted into a valid merge tree by recursively reverting the normalization,
thereby guaranteeing the validity of the interpolated BDT and merge tree.

%% file: notations.tex

\newcommand{\domain}{\mathcal{M}}
\newcommand{\range}{\mathbb{R}}
\newcommand{\sublevelset}[1]{#1^{-1}_{-\infty}}
\newcommand{\superlevelset}[1]{#1^{-1}_{+\infty}}
\newcommand{\Star}{St}
\newcommand{\Link}{Lk}
\newcommand{\simplex}{\sigma}
\newcommand{\face}{\tau}
\newcommand{\lowerlink}{\Link^{-}}
\newcommand{\upperlink}{\Link^{+}}
\newcommand{\Index}{\mathcal{I}}
\newcommand{\offset}{o}
\newcommand{\Natural}{\mathbb{N}}
\newcommand{\criticalSet}{\mathcal{C}}
\newcommand{\diagram}{\mathcal{D}}
\newcommand{\wasserstein}[1]{W^{\diagram}_#1}
\newcommand{\projection}{\Delta}
\newcommand{\hierarchy}{\mathcal{H}}
\newcommand{\decimation}{D}
\newcommand{\xDimD}{L_x^\decimation}
\newcommand{\yDimD}{L_y^\decimation}
\newcommand{\zDimD}{L_z^\decimation}
\newcommand{\xDim}{L_x}
\newcommand{\yDim}{L_y}
\newcommand{\zDim}{L_z}
\newcommand{\Grid}{\mathcal{G}}
\newcommand{\GridD}{\mathcal{G}^\decimation}
\newcommand{\x}{\phantom{x}}
\newcommand{\Mod}{\;\mathrm{mod}\;}
\newcommand{\NN}{\mathbb{N}}
\newcommand{\forwardIntegralLine}{\mathcal{L}^+}
\newcommand{\backwardIntegralLine}{\mathcal{L}^-}
\newcommand{\triangulationOp}{\phi}
\newcommand{\decimationOp}{\Pi}
\newcommand{\isovalue}{w}
\newcommand{\persistence}{\mathcal{P}}
\newcommand{\pointMetric}{d}
\newcommand{\diagramSet}{\mathcal{S}_\mathcal{D}}
\newcommand{\diagramSpace}{\mathbb{D}}
\newcommand{\jointree}{\mathcal{T}^-}
\newcommand{\splittree}{\mathcal{T}^+}
\newcommand{\mergetree}{\mathcal{T}}
\newcommand{\mergetreeSet}{\mathcal{S}_\mathcal{T}}
\newcommand{\branchset}{\mathcal{S}_\mathcal{B}}
\newcommand{\branchspace}{\mathbb{B}}
\newcommand{\mergetreeSpace}{\mathbb{T}}
\newcommand{\editdistance}{D_E}
\newcommand{\wassersteinTree}{W^{\mergetree}_2}
\newcommand{\distanceSequence}{d_S}
\newcommand{\branchtree}{\mathcal{B}}
\newcommand{\branchtreeSet}{\mathcal{S}_\mathcal{B}}
\newcommand{\branchtreeSpace}{\mathbb{B}}
\newcommand{\forest}{\mathcal{F}}
\newcommand{\sequenceSpace}{\mathbb{S}}
\newcommand{\forestMatrix}{\mathbb{F}}
\newcommand{\treeMatrix}{\mathbb{T}}
\newcommand{\normalizedLocation}{\mathcal{N}}
\newcommand{\normalizedWasserstein}{W^{\normalizedLocation}_2}
\newcommand{\geodesictree}{\mathcal{G}}
\newcommand{\dummyVector}{\mathcal{V}}
\newcommand{\geodesictreeVec}{g}
\newcommand{\geodesicAxis}{\mathcal{A}}
\newcommand{\directionVector}{\mathcal{V}}
\newcommand{\geodesicdiagram}{\mathcal{G}^{\diagram}}
\newcommand{\reconstructionError}{E_{L_2}}
\newcommand{\pcaBasis}{B_{\mathbb{R}^d}}
\renewcommand{\pcaBasis}{B}
\newcommand{\origin}{o_b}
\newcommand{\sizeEncoding}{n_e}
\newcommand{\sizeDecoding}{n_d}
\newcommand{\linearTransformation}{\psi}
\newcommand{\unitTransformation}{\Psi}
\renewcommand{\origin}{o}
\newcommand{\bdtOrigin}{\mathcal{O}}
\newcommand{\activation}{\sigma}
\newcommand{\validBDT}{\gamma}
\newcommand{\mtPgaBasis}{B_{\branchtreeSpace}}
\newcommand{\mtPgaError}{E_{\wassersteinTree}}
\newcommand{\frechetEnergy}{E_F}
\newcommand{\geodesicExtremity}{\mathcal{E}}
\newcommand{\vectorNotation}[1]{\protect\vv{#1}}
\renewcommand{\vectorNotation}[1]{#1}
\newcommand{\axisNotation}[1]{\protect\overleftrightarrow{#1}}
\newcommand{\individualEnergy}{E}
\newcommand{\ensembleSize}{N}
\newcommand{\numberBranchinBarycenter}{N_1}
\newcommand{\numberGeodesicSamples}{N_2}
\newcommand{\planarGridX}{N_x}
\newcommand{\planarGridY}{N_y}
\newcommand{\regularGrid}{G}
\newcommand{\distanceMatrix}{\mathbb{D}}
\newcommand{\maxDimensions}{{d_{max}}}
\newcommand{\projectionOperator}{\mathcal{P}}
\newcommand{\reconstructed}[1]{\widehat{#1}}
\newcommand{\gt}{>}
\newcommand{\lt}{<}
\newcommand{\branch}{b}
\newcommand{\nonLinearFunction}{\sigma}
\newcommand{\batchSequence}{S}

\newcommand{\julien}[1]{\textcolor{red}{#1}}
\newcommand{\mathieu}[1]{\textcolor{green}{#1}}
\renewcommand{\mathieu}[1]{\textcolor{green}{#1}}
\newcommand{\jules}[1]{\textcolor{orange}{#1}}
\renewcommand{\jules}[1]{\textcolor{black}{#1}}
\newcommand{\note}[1]{\textcolor{magenta}{#1}}
\newcommand{\cutout}[1]{\textcolor{blue}{#1}}
\renewcommand{\cutout}[1]{}

\newcommand{\revision}[1]{\textcolor{black}{#1}}
\newcommand{\minorRevision}[1]{\textcolor{black}{#1}}

\newcommand{\eqSpace}{-1.75ex}

\newcommand{\mycaption}[1]{
\caption{#1}
}